\documentclass[onecolumn]{aastex63}
\usepackage{amssymb}	
\usepackage{amsmath}	
\usepackage{array}
\usepackage{bm}
\usepackage{CJK}
\usepackage{graphicx}	
\usepackage{hyperref}
\usepackage{xifthen}
\usepackage{xspace}
\usepackage{natbib}
\newcommand{\Proj}{\mathcal{P}}
\newcommand{\Deproj}{\mathcal{P}^{-1}}
\newcommand{\aizw}[1]{#1}
\newcommand{\aizwt}[1]{#1}

\begin{document}
\begin{CJK*}{UTF8}{gbsn}
\title{On the Interpretation of Velocity Residuals in Protoplanetary Disks }

\author[0000-0001-8877-4497]{Masataka Aizawa}
\affiliation{College of Science, Ibaraki University, 2-1-1 Bunkyo, Mito, 310-8512, Ibaraki, Japan }

\author[0000-0003-4039-8933]{Ryuta Orihara}
\affiliation{Department of Astronomy, Graduate School of Science, The University of Tokyo, 7-3-1 Hongo, Bunkyo-ku, Tokyo 113-0033, Japan}

\begin{abstract}
We present a first-order analytical model for line-of-sight velocity residuals, defined as the difference between observed velocities and those predicted by a fiducial model, assuming a flared, nearly axisymmetric disk with the perturbations in disk surface height $\delta h(r)$, inclination $\delta i(r)$, and position angle $\delta\mathrm{PA}(r)$.  Introducing projection-deprojection mapping between sky-plane and disk-frame coordinates, we demonstrate that the normalized velocity residuals exhibit Fourier components up to the third harmonic ($\sin3\phi$ and $\cos3\phi$). Moreover, we show that the radial profiles of $\delta h(r)$, $\delta i(r)$, and $\delta\mathrm{PA}(r)$ can be uniquely recovered from the data by solving a linear inverse problem.  For comparison, we highlight factors that are not considered in previous models. We also outline how our framework can be extended beyond the first-order residuals and applied to additional observables, such as line intensities and widths.
\end{abstract}


\section{Introduction}
Recent ALMA observations of protoplanetary disks have revealed highly structured gas disks traced in multiple molecular lines \citep[e.g.,][]{oberg2021}. Departures from pure Keplerian rotation could signal substructures, such as warps \citep[e.g.,][]{casassus2015} or the presence of embedded planets \citep[e.g.,][]{pinte2018,pinte2020}. In particular, the exoALMA Large Program provides unprecedented data for studying the kinematic structure of disks \citep{teague2025}. The survey demonstrates that disk velocity fields often display large-scale asymmetries beyond simple Keplerian profiles \citep{izquierdo2025,stadler2025}. Interpreting this wealth of information is essential for advancing our understanding of planet formation.

\cite{winter2025} present an analytical model of a slightly warped disk and apply it to the exoALMA observations to explain the observed large-scale velocity residuals. In their approach, they introduce warp structures via small, radial variations in  the position angle $\delta \mathrm{PA}$ and inclination $\delta i$ of disk, and show that some of the observed residuals are consistent with these perturbations. Indeed, numerous protoplanetary disks show warp signatures in velocity maps \citep[e.g.,][]{casassus2015} and in scattered-light shadows \citep[e.g.,][]{marino2015,benisty2017}, although radial flows can complicate the interpretation of twisted velocity patterns for the former case \citep{rosenfeld2014,zuleta2024}. 

However, two factors not considered in \cite{winter2025} can substantially affect the interpretation of the observed velocity residuals. First, a rigorous treatment of the projection-deprojection mapping between sky-plane and disk-plane coordinates is essential when modeling observed velocities \cite[e.g.,][]{casassus2019}. Second, the disk surface height profile, $h(r)$, and its deviation from the fiducial model, $\delta h(r)$, influence both the velocity field and the projection-deprojection mapping. Indeed, ALMA observations of disk surface emission have revealed vertical substructures \citep[e.g.,][]{pinte2018_2,law2021,law2022,panequecarreno2023}, indicating that true height profiles often deviate from the simple prescriptions commonly adopted in disk modeling \citep[e.g.,][]{izquierdo2021}. In Section \ref{sec:comment_winter}, we summarize the factors that the models presented in \cite{winter2025} do not account for. 

Previously, \cite{casassus2019} present a numerical methodology for inferring radial variations in disk structures, namely position angle, inclination, and surface height, from observed velocity fields. In this paper, we present a first-order analytical framework that computes line-of-sight velocity residuals in flared, nearly axisymmetric disks, considering the projection-deprojection mapping. Using the framework, we also present a method for recovering disk structural perturbations from observational data. 

The paper is organized as follows. In Section~\ref{sec:residual_general}, we present a general formulation for deriving velocity residuals. Section~\ref{sec:first_order} develops the first-order expansion of these residuals. In Section \ref{sec:determine_perturb}, we discuss how we can recover  $\delta h(r), \delta i(r), \delta\mathrm{PA}(r)$  from the observation. Finally, Section~\ref{sec:conclusions} summarizes our main conclusions and discusses future prospects.

\section{Formulation for velocity residuals } \label{sec:residual_general}

\subsection{Coordinate and line-of-sight velocity} \label{sec:coordinate}
\begin{figure*}
\centering
\includegraphics[width=0.7\linewidth]{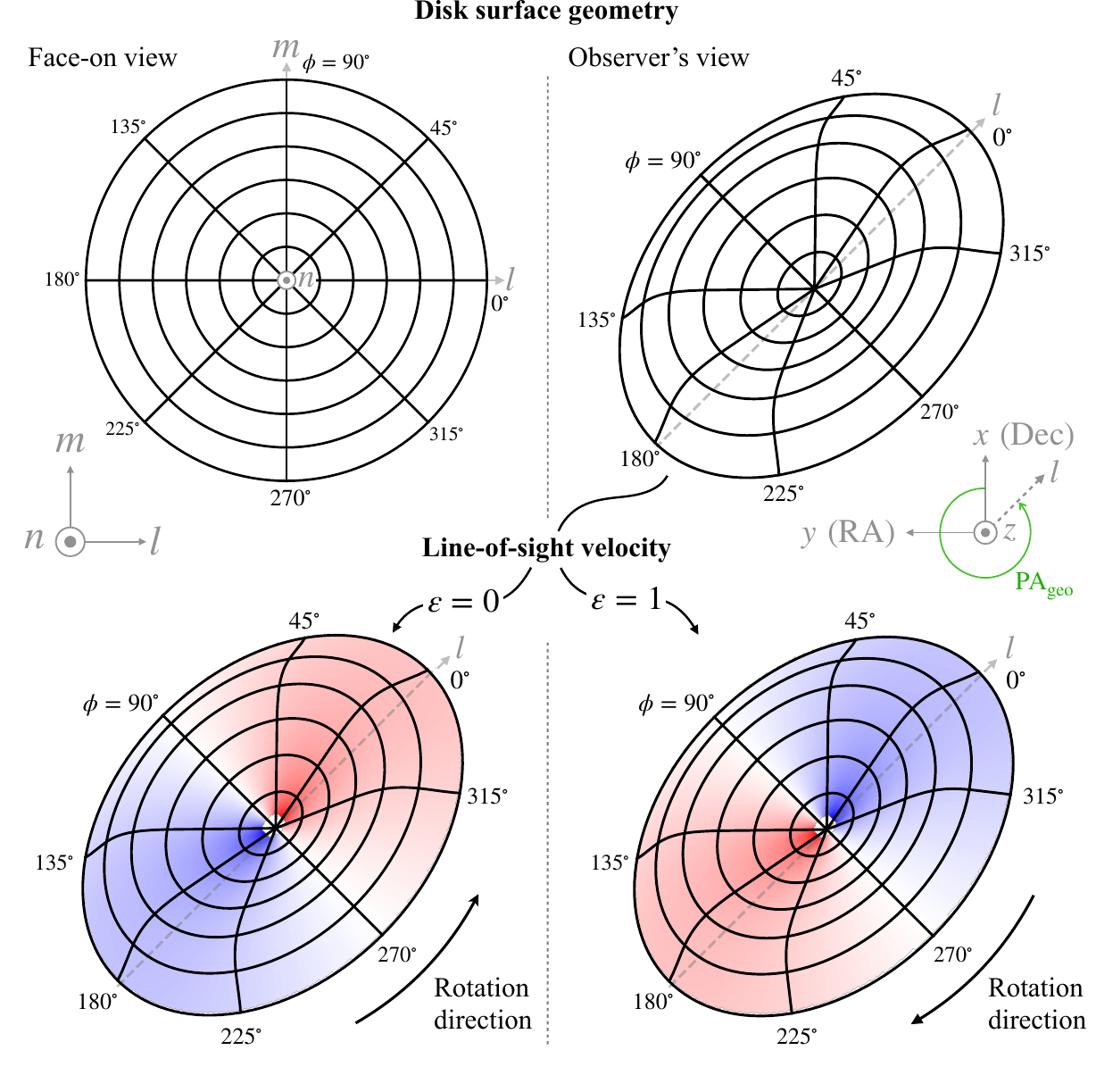}
\caption{ \aizwt{Illustration of the coordinate system used in our model, as described in Section~\ref{sec:coordinate}. The upper panels show face-on and observer's views of the disk surfaces, along with both the disk-plane coordinates $(l,m,n)$ and the sky-plane coordinates $(x,y,z)$. The lower panels present velocity fields for counterclockwise ($\epsilon=0$) and clockwise ($\epsilon=1$) rotations. We adopt $\mathrm{PA}_{\rm geo} =  7 \pi /4$. }
}
\label{fig:geo_vel}
\end{figure*}

\aizwt{To consider velocity fields on the disk surface, we} define coordinate frames for the sky and the disk surface following \cite{orihara2025}.  \aizwt{We summarize the coordinate system with velocity fields in Figure \ref{fig:geo_vel}. We begin by explicitly defining the disk-surface coordinate system, and then map the velocity fields onto the disk surface. } 

For the sky coordinate, we adopt a basis $\{\hat{\bm{e}}_x,\hat{\bm{e}}_y,\hat{\bm{e}}_z\}$ defined on the celestial sphere. In this frame,  $\hat{\bm{e}}_x$ points toward increasing declination (i.e., northward on the sky),  $\hat{\bm{e}}_y$ points toward increasing right ascension  (i.e.\ eastward on the sky),  and $\hat{\bm{e}}_z$ is aligned with the line of sight toward the observer (from the source to the telescope). In this sky frame, we introduce a local disk basis  $\{\hat{\bm{l}},\hat{\bm{m}},\hat{\bm{n}}\}$, defined by the disk's position angle, \aizwt{$0 \leq \mathrm{PA}_{\mathrm{geo}} < 2\pi$, which is newly introduced to clarify the near-far correspondence of the disk, and inclination $0 \leq i \leq \pi/2$}:
\begin{align}
\hat{\bm{l}} &= (\cos \mathrm{PA}_{\mathrm{geo}},\sin \mathrm{PA}_{\mathrm{geo}},0) \\
\hat{\bm{m}} &= (-\cos i \sin \mathrm{PA}_{ \mathrm{ geo}},\cos i \cos \mathrm{PA}_{\mathrm{geo}},-\sin i) \\
\hat{\bm{n}} &= (-\sin i \sin \mathrm{PA}_{\mathrm{geo}},\sin i \cos \mathrm{PA}_{ \mathrm{geo}},\cos i), 
\end{align}
where $\hat{\bm{l}}$ points along the disk major axis, $\hat{\bm{n}}$ is the unit vector normal to the disk plane (i.e., the disk axis), and $\hat{\bm{m}}$ is defined to complete the left-handed orthonormal basis. \aizwt{Here, we define ``the geometric position angle", $\mathrm{PA}_{\mathrm{geo}}$, such that the direction obtained by rotating $\hat{\bm{l}}$ counterclockwise by $\pi/2$ on the sky corresponds to the far side of the upper surface of the disk. This definition differs from the kinematic position angle, which we refer to as $\mathrm{PA}_{{\mathrm{red}}}$, measured from north to the redshifted side of the disk major axis, as is typically determined from observations of velocity fields. Although $\mathrm{PA}_{{\mathrm{red}}}$ is commonly adopted for modeling of velocity fields, we adopt the geometric definition  for PA throughout the paper to clarify the coordinate of the disk surface, 
\begin{equation}
    \mathrm{PA} = \mathrm{PA}_{\rm geo}. 
\end{equation} }

A point on the disk surface $\mathbf S$ at cylindrical coordinates $(r,\phi)$ and height $h(r)$ has Cartesian coordinate
\begin{align}
\mathbf S &= r (\cos\phi \hat{\bm{l}} + \sin\phi \hat{\bm{m}} ) + \aizw{h(r)} \hat{\bm{n}},    \\ 
&=\begin{pmatrix}
r (\cos\phi\cos \mathrm{PA} - \cos i \sin\phi \sin \mathrm{PA} ) - h(r) \sin i \sin \mathrm{PA}\\
r (\cos\phi\sin \mathrm{PA} + \cos i \sin\phi \cos \mathrm{PA} ) + h(r) \sin i \cos \mathrm{PA}\\
-r \sin i \sin\phi + h(r) \cos i
\end{pmatrix},  \label{eq:matrix_S}
\end{align}
where only the upper surface is considered. Here we measure the azimuthal angle $\phi$ from the major axis $\phi=0$ in the disk plane: $\phi=\pi/2$ corresponds to the far side of the upper surface (farthest from the observer), while $\phi=3\pi/2$ corresponds to the near side. The unit vector in the azimuthal direction is
\begin{equation}
 \hat{\bm{e}}_{\phi} =  (-\sin \phi \hat{\bm{l}}  + \cos\phi \hat{\bm{m}} ), 
\end{equation}

\aizwt{For a stellar mass of $M_{\star}$, the Keplerian rotational velocity for a geometrically flared disk at radius $r$ and height $h(r)$ on the upper surface is given by \cite[e.g.,][]{rosenfeld2013} 
\begin{equation}
  v_\phi(r)
  = \frac{\sqrt{GM_\star}r}{(r^2 + h(r)^2)^{3/4}}, \label{eq:vlos}
\end{equation}
where $G$ is the gravitational constant. We here explicitly neglect both radial pressure gradients and disk self-gravity to isolate the purely geometric effect of the emission surface height. Deviations due to introduced by these processes are examined in previous studies (e.g., \cite{stadler2025} and \cite{longarini2025}, for the exoALMA data.).}

Assuming all emission arises from this upper layer, the line-of-sight velocity is then 
\begin{align}
v_{\rm los}(r,\phi) &=  (-1)^{\varepsilon} (- v_\phi(r)  (\hat{\bm{e}}_{\phi}  \cdot  \hat{\bm{e}}_z))\\ &= (-1)^{\varepsilon}  \frac{\sqrt{GM_\star}r}{(r^2 + h(r)^2)^{3/4}}\sin i \cos \phi \label{eq:los_vel}
\end{align}
where the direction of the disk rotation is specified by $\varepsilon$, \aizw{in the same manner as a sign function in \cite{izquierdo2025}}: 
\begin{align}
    \varepsilon = 
\begin{cases}
0, & \text{if rotation is counterclockwise to the observer}, \\
1, & \text{if rotation is clockwise to the observer}, 
\end{cases}
\end{align}
When $\varepsilon=1$, $\phi=0$ corresponds to the blue-shifted direction, whereas when $\varepsilon=0$, $\phi=0$ corresponds to the red-shifted direction $\varepsilon$. \aizwt{In this case, the relation between $\mathrm{PA}_{\rm geo}$ and $\mathrm{PA}_{\rm red}$ is given as follows: 
\begin{align}
\mathrm{PA}_{\rm geo} = 
\begin{cases}
\mathrm{PA}_{\rm red} , & \text{if $\varepsilon = 0$}, \\
\mathrm{PA}_{\rm red} + \pi , & \text{if $\varepsilon = 1$}. 
\end{cases}
\end{align}
}

\aizwt{Our defined coordinate system differs from those adopted in previous studies. Some works define the azimuthal angle such that $\phi=0$ points to the redshifted side \citep{rosenfeld2013,casassus2019, casassus2022}. To account for both of rotational directions, these studies allow the inclinations to extend beyond  $\pi/2$, such that for $\pi/2 \leq i$, the lower-surface of the disk faces the observer and the apparent sense of rotation is reverse, expressed as retrograde motion. Other works instead adopt negative inclinations with $-\pi/2 \leq i \leq \pi/2$ \citep{pinte2018,teague2019,izquierdo2025}, in which case $\phi=0$ correspond to the blueshifted direction for $i<0$. In \cite{izquierdo2025}, a rotational direction is specified by an additional parameter equivalent to $\varepsilon$\footnote{\citet{izquierdo2025} adopt models with $-\pi/2 < i < \pi/2$, introducing a rotational-direction parameter $\text{sgn}_{\rm rot}$, equivalent to $\epsilon$, as listed in their Table~2. However, $\text{sgn}_{\rm rot}$ becomes redundant once $-\pi/2 < i < \pi/2$ is specified.}. }

\aizwt{Since this study examines the coordinate transformation between the sky and disk planes, an unambiguous definition of the surface coordinate system is essential for clarity. Therefore, we adopt explicitly define the coordinate system for the disk surface with $0 \leq i \leq \pi/2$, clarifying near-far correspondence; $\phi=\pi/2$ and $3 \pi/2$ correspond to the far and near sides of the upper disk surfaces, respectively, regardless of the rotational direction. The rotational direction is then separately introduced through the parameter $\varepsilon$. Note that this coordinate definition, independent of the velocity field, is also advantageous for considering other physical quantities (e.g., intensities) in our perturbative framework in future work. }

\subsection{True and approximate models}
To interpret the observed velocity residuals, we introduce two disk-geometry models:

\begin{description}
  \item[Model 0 (approximate)] 
 The fiducial geometry used to fit the data,
 \begin{align}
      \{h_{0}(r),i_{0},\mathrm{PA}_{0}\}.
    \end{align}
  \item[Model 1 (true)] 
 The actual disk geometry,
 \begin{align}
      \{h(r),i(r),\mathrm{PA}(r)\}
      = \bigl\{h_{0}(r)+\delta h(r),i_{0}+\delta i (r),\mathrm{PA}_{0}+\delta\mathrm{PA}(r)\}.
    \end{align}
\end{description}
Unless otherwise stated, in the majority of expressions we drop the $(r)$ argument and denote  
\begin{equation}
    (\delta h(r),\,\delta i(r),\,\delta\mathrm{PA}(r)) \to  (\delta h,\,\delta i,\,\delta\mathrm{PA}). 
\end{equation}

The measured velocity field follows Model 1, whereas Model 0 represents the best-fit geometry. We define 
\begin{align}
v_{\rm los, 0} (r, \phi) &\equiv v_{\rm los} (r, \phi |  h_0(r) , \mathrm{PA}_0 ,i_0 ),  \\
v_{\rm los, \delta } (r, \phi) &\equiv v_{\rm los} (r, \phi |  h_0(r) + \delta h, i_0 + \delta i, \mathrm{PA}_0 + \delta {\rm \mathrm{PA}}).
\end{align}

\subsection{Deprojection/Projection mapping} \label{sec:depro_pro}
We define projection and deprojection mapping between the sky-plane coordinates $(x,y)$ and the disk-plane coordinates $(r,\phi)$: 
\begin{align}
  (x, y) &= \Proj_{h(r), i, \mathrm{PA}}(r,\phi)
  \quad\text{(projection onto the sky plane),} \\
  (r, \phi) &= \Deproj_{h(r),i,  \mathrm{PA}}(x,y)
  \quad\text{(deprojection onto the disk plane).} 
\end{align}
These projection and deprojection operators are conceptually equivalent to $\bm{f}_{\mathrm{PA}, i, +\psi}$ and $\bm{f}^{-1}_{\mathrm{PA}, i, +\psi}$ in \cite{casassus2019} where $\psi$ denotes the opening angle function for the disk, although the definitions of coordinates, functions, and parameters differ.

For a highly inclined disk with a large height function, the projection mapping $\Proj$ can be non-injective, so that distinct $(r,\phi)$ may map to the same sky-plane coordinate $(x,y)$, in contrast to \aizwt{the case of the conical surface considered in} \cite{casassus2019}. In other words, the deprojection mapping $\Deproj$ can be multi-valued: $\Deproj$ may yield multiple $(r,\phi)$ solutions. Nevertheless, for simplicity we assume that $\Proj$ is injective, so that each $(x,y)$ corresponds uniquely to a single $(r,\phi)$.

From Equation (\ref{eq:matrix_S}), the projection onto the sky plane is
 \begin{align}
x &= r (\cos\phi\cos\mathrm{PA} - \cos i \sin\phi \sin\mathrm{PA}) - h(r) \sin i \sin\mathrm{PA}, \\
y &= r(\cos\phi\sin\mathrm{PA} + \cos i \sin\phi \cos\mathrm{PA}) + h(r) \sin i \cos\mathrm{PA}.
\end{align}
 Since the projection depends on the assumed geometry, we introduce
 \begin{align}
\Proj_0 &\equiv \Proj_{h_0(r),i_0, \mathrm{PA}_0},  \\
\Proj_\delta &\equiv \Proj_{h_0(r) + \delta h, i_0 + \delta i, \mathrm{PA}_0 + \delta {\rm \mathrm{PA}}}, 
\end{align}
for Model 0 (approximate) and Model 1 (true), respectively.

\subsection{Velocity residuals in observation}
\label{sec:vel_res_obs}
In the ideal (noise-free) case without beam smearing, the measured velocity field $v_{\rm obs}(x,y)$ equals the true line-of-sight velocity evaluated with the true disk coordinate, 
\begin{equation}
    v_{\rm obs}(x,y)= v_{\rm los, \delta }(r, \phi), \quad  (x,y)= \Proj_{\delta} (r, \phi). \label{eq:vel_obs_sky}
\end{equation}
Here $v_{\rm los, \delta }(r, \phi)$ and $\Proj_{\delta}$ refer to the “true” geometry (Model 1). Note that the disk coordinate $(r, \phi)$ themselves depend on $\Proj_{\delta}$; observers only know $v_{\rm obs}(x,y)$ and must infer $\Proj_{\delta}$.

By contrast, the modeled velocity field uses the fiducial (Model 0) geometry:
\begin{equation}
    v_{\rm Model \, 0} (x,y)= v_{\rm los, 0}(r, \phi), \quad  (x,y)= \Proj_{0} (r, \phi). 
\end{equation}
Thus, the velocity residual in the sky plane $\delta v_{\rm obs}(x, y) $ resulting from fitting the model to the data can be expressed as: 
\begin{align}
     \delta v_{\rm obs}(x, y) &=v_{\rm obs}(x,y) -  v_{\rm Model \, 0} (x,y) \\
    &= v_{\rm los, \delta }(\Proj_{\delta}^{-1}(x,y)) - v_{\rm los, 0 }(\Proj_{0}^{-1}(x,y)) \\
    &= [v_{\rm los, \delta }(\Proj_{\delta}^{-1}(x,y)) - v_{\rm los, \delta }(\Proj_{0}^{-1}(x,y))] +[v_{\rm los, \delta }(\Proj_{0}^{-1}(x,y)) - v_{\rm los, 0 }(\Proj_{0}^{-1}(x,y))].  \label{eq:project_v_res}
\end{align}
Here, the first term represents the error introduced by deprojection (mapping error), while the second term represents the difference in the functional form of $ v_{\rm los}$. 

In the realistic situation, one deprojects the observed velocity using the fiducial model:
\begin{equation}
    v_{\rm obs, depro}(r,\phi)= v_{\rm obs}(x,y), \quad (r,\phi) = \Proj_0^{-1} (x,y), 
\end{equation}
Using Equation (\ref{eq:vel_obs_sky}) we  obtain
\begin{equation}
    v_{\rm obs, depro}(r,\phi)= v_{\rm los, \delta }( \Proj_{\delta}^{-1} ( \Proj_0 (r,\phi)) ).
\end{equation}
 Similarly, the deprojected fiducial model is
 \begin{equation}
    v_{\rm 0, depro}(r,\phi) = v_{\rm los, 0 }( \Proj^{-1}_{0} (\Proj_{0}(r, \phi))) =  v_{\rm los, 0 }(r, \phi). 
\end{equation}
Therefore, the residual in disk coordinates is
\begin{align}
      \delta v_{\rm obs, depro}(r,\phi)
  &=  v_{\rm obs, depro}(r,\phi) -  v_{\rm 0, depro}(r,\phi) \\
  &= v_{\rm los, \delta }( \Proj_{\delta}^{-1} ( \Proj_0 (r,\phi))) - v_{\rm los, 0 }(r, \phi) \\
  &= \underbrace{[v_{\rm los, \delta }(  \Proj_{\delta}^{-1} ( \Proj_0 (r,\phi))) - v_{\rm los, \delta }(r, \phi)]}_{\text{mapping error}} + \underbrace{[v_{\rm los, \delta }(r, \phi)  - v_{\rm los, 0 }(r, \phi)]}_{\text{model difference}}.\label{eq:velocity_residual}
\end{align}
Similar to Equation (\ref{eq:project_v_res}), the first term corresponds to the (projection-deprojection) mapping error, while the second term corresponds to the functional difference in $v_{\rm los}$.

In the next section we expand Equation (\ref{eq:velocity_residual}) to first order in the small perturbations $(\delta h,\delta i,\delta\mathrm{PA})$.  The deprojected velocity residual
$\delta v_{\rm res,\,depro}(r,\phi)$ can be mapped back to sky-plane coordinates simply by applying the fiducial projection operator $\Proj_{0}(r,\phi)$. Since $\delta v_{\rm res,\,depro}$ is already $\mathcal{O}(\delta)$, any error from using $\Proj_{0}$ instead of the exact projection is only $\mathcal{O}(\delta^2)$ and can be neglected.

\section{First-order expansion of velocity residual} \label{sec:first_order} 
\subsection{Mapping-error and model-difference terms}\label{sec:first_order}
We expand Equation~(\ref{eq:velocity_residual}) to first order in the perturbations $(\delta h,\delta i,\delta\mathrm{PA})$.  As noted above, the total residual in Equation~(\ref{eq:velocity_residual}) naturally decomposes into two contributions: a mapping-error term, which captures projection-deprojection mismatches, and the model-difference term. The model-difference contribution is then given by:
\begin{align}
 v_{\rm los, \delta }(r, \phi)  - v_{\rm los, 0 }(r, \phi) &= \left(\frac{\partial{ v_{\rm los}}}{{\partial i}}\right) \delta i + \left(\frac{\partial{ v_{\rm los}}}{{\partial h}}\right) \delta h + \left(\frac{\partial{ v_{\rm los}}}{{\partial \mathrm{PA}}}\right) \delta \mathrm{PA}+ \mathcal{O}(\delta^2) \label{eq:model_dif_0} \\
 &= (-1)^{\varepsilon}[V_{K,0}(r) \cot i_0 \cos\phi] \delta i - (-1)^{\varepsilon}\left[V_{K,0}(r) \left(\frac{3h_0(r) }{2(r^2 + h_0^2(r))}\right) \cos\phi \right] \delta h + \mathcal{O}(\delta^2),  \label{eq:model_dif}
\end{align}
where we define
\begin{equation}
    V_{K,0}(r) =\frac{\sqrt{GM_\star}r}{(r^2 + h_0(r)^2)^{3/4}}
    \sin i_0. 
\end{equation}

To compute the mapping error, we define small shits $\delta r$ and $\delta \phi$ by
\begin{align}
 \Proj_{\delta}^{-1} ( \Proj_0 (r,\phi)))  = (r+\delta r, \phi + \delta \phi). \label{eq:dr_dphi}
\end{align}
Then, the mapping error can be computed as: 
\begin{align}
v_{\rm los, \delta }( \Proj_{\delta}^{-1} ( \Proj_0 (r,\phi))) - v_{\rm los, \delta }(r, \phi)
&= v_{\rm los, \delta } (r+\delta r, \phi + \delta \phi) - v_{\rm los, \delta }(r, \phi) \\
&= \left(\frac{\partial{v_{\rm los, \delta }}}{{\partial r}}\right) \delta r+ \left(\frac{\partial{v_{\rm los, \delta }}}{\partial \phi}\right) \delta \phi + \mathcal{O}(\delta^2) \\
&= \left(\frac{\partial{v_{\rm los, 0 }}}{{\partial r}}\right) \delta r+ \left(\frac{\partial{v_{\rm los, 0 }}}{\partial \phi}\right) \delta \phi+ \mathcal{O}(\delta^2), \label{eq:project_deproject_eq}
\end{align}
where one finds 
\begin{align}
\frac{\partial{v_{\rm los, 0 }}}{{\partial r}} 
&= - (-1)^{\varepsilon} V_{K,0}(r) A(r) \cos \phi, \\
\frac{\partial{v_{\rm los, 0 }}}{\partial \phi}
&= - (-1)^{\varepsilon} V_{K,0}(r) \sin\phi, \\
A(r) &\equiv   \left(\frac{r + 3 h_0(r) h_0'(r) - 2 h_0^2(r)/r}{2(r^2 + h_0^2(r))} \right).
\end{align}

Adding Equations (\ref{eq:model_dif}) and (\ref{eq:project_deproject_eq}) gives the full first-order expansion of Equation~(\ref{eq:velocity_residual}). 

\subsection{Calculation of $\delta r, \delta \phi $ due to mapping error} \label{sec:dr_dphi_map_error}
To compute $(\delta r, \delta \phi)$ in Equation (\ref{eq:dr_dphi}), induced by perturbation in $(\delta h,\delta i,\delta\mathrm{PA})$, we define the projection functions
\begin{align}
f(r, \phi ,h(r), \mathrm{PA}, i) & \equiv x = r (\cos\phi\cos\mathrm{PA} - \cos i \sin\phi \sin\mathrm{PA}) - h(r) \sin i \sin\mathrm{PA},  \label{eq:df} \\
 g(r, \phi , h(r), \mathrm{PA}, i)  &\equiv  y =r(\cos\phi\sin\mathrm{PA} + \cos i \sin\phi \cos\mathrm{PA}) + h(r) \sin i \cos\mathrm{PA}. \label{eq:dg} 
\end{align}

Taking the total differentials of $f$ and $g$, we find: 
\begin{align}
df &= f_r \delta r + f_\phi \delta \phi + f_h \delta h  + f_i \delta i  + f_{\mathrm{PA}} \delta \mathrm{PA}, \\ 
dg &= g_r \delta r + g_\phi \delta \phi + g_h \delta h  + g_i \delta i  + g_{\mathrm{PA}} \delta \mathrm{PA}.
\end{align}
\aizwt{
where subscripts denote partial derivatives, for example, 
\begin{equation}
    f_r\equiv\frac{\partial f}{\partial r}. 
\end{equation}
}

Fixing $(x,y)$ on the sky coordinate, we obtain 
\begin{equation}
   df= dx=0, \quad  dg=dy=0.  \label{eq:df_dg_0}
\end{equation} 
Combining Equation (\ref{eq:df}), (\ref{eq:dg}), and (\ref{eq:df_dg_0}), we find 
\begin{align}
\begin{pmatrix}
\delta r \\[3pt]
\delta\phi
\end{pmatrix}
=
-\frac{1}{f_r g_\phi - f_\phi g_r }
\begin{pmatrix}
g_\phi & -f_\phi \\
-g_r & f_r
\end{pmatrix}
\begin{pmatrix}
f_h & f_i & f_{\mathrm{PA}} \\
g_h & g_i & g_{\mathrm{PA}}
\end{pmatrix}
\begin{pmatrix}
\delta h \\
\delta i \\
\delta\mathrm{PA}
\end{pmatrix}, \label{eq:dr_dphi}
\end{align}
Equivalently, we have \aizwt{
\begin{align}
\delta r &=r_h \delta h+ r_i \delta i+ r_{\rm PA} \delta \mathrm{PA}, \\
\delta \phi &= \phi_h \delta h+ \phi_i \delta i+ \phi_{\rm PA} \delta \mathrm{PA}, \\
r_p &=\frac{ f_\phi g_p - g_\phi f_p}{f_r g_\phi - f_\phi g_r},\quad
\phi_p =\frac{ g_r f_p - f_r g_p }{f_r g_\phi - f_\phi g_r}
\quad (p = h,\mathrm{PA},i ). \label{eq:partial_deri_eq}
\end{align}}
In Appendix (\ref{sec:3_detailed_dr_dphi}), we show derivations of $f_r, f_{\phi}, g_r, g_{\phi}, r_{p}, \phi_p$.

For convenience, we define the denominator for $r_p$ and $\phi$ in Equation (\ref{eq:partial_deri_eq}) as  
\begin{equation}
    \Delta = \Delta(r,\phi) \equiv  f_r g_\phi - f_\phi g_r = r(\cos i_0+ h_0'(r)\sin i_0\sin\phi). 
\end{equation}
The mapping $(r,\phi)\to(x,y)$ becomes singular when $\Delta=0$, i.e.\ when
\begin{align}
 h_0'(r)\sin i_0\sin\phi  =-\cos i_0,  
\end{align}
which corresponds to the line of sight being tangent to the flared surface (the ``rim").  Although this can occur for inclined or strongly flared disks, for each $r$ there are at most two solutions in $\phi\in[0,2\pi)$, so the singular locus has measure zero and may be ignored.  We therefore assume $\Delta\neq0$ in what follows.

\subsection{First-order expansion of velocity residuals}
Using Equation~(\ref{eq:dr_dphi}), the deprojection (mapping)-error term in Equation~(\ref{eq:project_deproject_eq}) can be written as
\begin{align}
&[v_{\rm los, \delta }(  \Proj_{\delta}^{-1} ( \Proj_0 (r,\phi))) - v_{\rm los, \delta }(r, \phi)] \\
&= (-1)^{\varepsilon+1}  \frac{V_{K,0}(r)}{\Delta(r,\phi)} \Delta \left[
  \left(\frac{\partial r}{\partial h}\delta h + \frac{\partial r}{\partial i}\delta i + \frac{\partial r}{\partial \mathrm{PA}_0}\delta\mathrm{PA}_0\right)
  A(r) \cos\phi
  + \left(\frac{\partial \phi}{\partial h}\delta h + \frac{\partial \phi}{\partial i}\delta i + \frac{\partial \phi}{\partial \mathrm{PA}_0}\delta\mathrm{PA}_0\right)\sin\phi
\right]  + \mathcal{O}(\delta^2) \\
&= (-1)^{\varepsilon+1} \frac{V_{K,0}(r)}{\Delta(r,\phi)} \left[B(r, \phi) \delta h  + C(r, \phi)\delta i  + D(r, \phi)\delta \mathrm{PA} \right] + \mathcal{O}(\delta^2),  \label{eq:v_deproject}
\end{align}
where we summarize the terms using
\begin{align}
B(r,\phi) &= B_0(r) \sin2\phi , \\
C(r,\phi) &= C_0(r) \cos\phi + C_1(r) \sin2\phi + C_2(r)\cos3\phi , \\
D(r,\phi) &= D_0(r)
           + D_1(r)\sin\phi
           + D_2(r)\cos2\phi
           + D_3(r)\sin3\phi.
\end{align}
The detailed functional forms for the coefficients $B_i(r), C_i(r), D_i(r)$ are presented in Appendix \ref{sec:3_depro_error}.  

Finally, combining Equations~(\ref{eq:model_dif}) and (\ref{eq:v_deproject}), we obtain the velocity residuals in Equation (\ref{eq:velocity_residual}):   
\begin{align}
&\delta v_{\rm obs, depro}(r,\phi)= [v_{\rm los, \delta }(  \Proj_{\delta}^{-1} ( \Proj_0 (r,\phi))) - v_{\rm los, \delta }(r, \phi)] + [v_{\rm los, \delta }(r, \phi)  - v_{\rm los, 0 }(r, \phi)] \\
&= (-1)^{\varepsilon+1}  \frac{V_{K,0}(r)}{\Delta(r,\phi)} \left[B(r, \phi) \delta h  + C(r, \phi)\delta i  + D(r, \phi)\delta \mathrm{PA}  - [\Delta(r,\phi) \cot i_0\cos \phi]  \delta i + \left[\Delta(r,\phi) \left(\frac{3h_0(r) }{\aizw{2}(r^2 + h_0^2(r))}\right) \cos \phi\right] \delta h\right] + \mathcal{O}(\delta^2), \label{eq:v_combined_raw}
\end{align}

We then decompose the last two terms as
\begin{align}
  -\Delta(r,\phi) \cot i_0\cos \phi &=  E_0(r) \cos \phi + E_1(r) \sin 2 \phi,  \\
\Delta(r,\phi) \left(\frac{3h_0(r) }{2(r^2 + h_0^2(r))}\right) \cos \phi&=    F_0(r) \cos \phi + F_1(r) \sin 2\phi. 
\end{align}
The detailed functional forms for $E_i(r), F_i(r)$ are presented in Appendix \ref{sec:3_depro_error}. 

Multiplying Equation~(\ref{eq:v_combined_raw}) by $(-1)^{\varepsilon+1}\Delta/V_{K,0}$ then gives
\begin{align}
(-1)^{\varepsilon+1} \frac{\delta v_{\rm res,depro}(r,\phi)\,\Delta(r,\phi)}{V_{K,0}(r)}
&=
\bigl[F_0(r)\cos\phi + \bigl(B_0(r)+F_1(r)\bigr)\sin2\phi\bigr]
\,\delta h
\nonumber\\
&\quad+
\bigl[(C_0(r)+E_0(r))\cos\phi
               + (C_1(r)+E_1(r))\sin2\phi
               + C_2(r)\cos3\phi\bigr]
\,\delta i
\nonumber\\
&\quad+
\bigl[D_0(r)
               + D_1(r)\sin\phi
               + D_2(r)\cos2\phi
               + D_3(r)\sin3\phi\bigr]
\,\delta\mathrm{PA}\, + \mathcal{O}(\delta^2).
\label{eq:delta_final_clean}
\end{align}
The equivalent Fourier-series expansion is given in Appendix~\ref{sec:3_fourier_exp}. 

In Equation~(\ref{eq:delta_final_clean}), the quantities on the left-hand side can be computed directly from the observational data and the fiducial model, while the coefficients on the right-hand side are determined entirely by the fiducial geometry.  Hence, one can invert Equation~(\ref{eq:delta_final_clean}) to recover $\delta h(r)$, $\delta i(r)$, and $\delta\mathrm{PA}(r)$, as described in Section~\ref{sec:determine_perturb}.

\subsection{Comparison of numerical calculation and analytical solution }

To validate our analytical model, we compare its predictions with numerically computed velocity fields, adopting the J1615 disk from the exoALMA survey \citep{teague2025} as our fiducial case. We set
\begin{align}
    M_{\star} = 1.14 M_{\odot}, \, 
    i_0 = 45^{\circ}, \, 
    \mathrm{PA}_0 =  3 \pi /2, \,
    \varepsilon = 1, 
\end{align}
where PA is taken to be $-90^{\circ}$ for visualization. Here, the west side is blue-shifted and the azimuthal angle $\phi$ is measured counter-clockwise from the blue-shifted major axis.  We assume a flared surface assuming radial range $r\in[30,600]$\,au: 
\begin{equation}
    h(r)=z_{0}\Bigl(\frac{r}{D_{0}}\Bigr)^{p}
              \exp\bigl[-(r/R_{t})^{q}\bigr] 
\end{equation}
with parameters for $^{12}\mathrm{CO}$ from \cite{izquierdo2025}:
\begin{align}
 z_{0}=26.3\,\mathrm{au}, \,
 D_{0}=100\,\mathrm{au}, \, p=1.04,  \,
 R_{t}=530\,\mathrm{au},  \,
 q=6.89. 
\end{align}
Its derivative is given by
\begin{equation}
h'(r) = h(r) \left[p/r -(q/R_t)(r/R_{t})^{q-1}  \right]. 
\end{equation}

We impose a perturbation
\begin{equation}
\delta h (r)=0.1 h(r), \, \delta i = 1^{\circ}, \, \delta \mathrm{PA}=
1^{\circ} 
\end{equation}
and compute two velocity fields for Model 0 (approximate) and Model 1 (true). \aizwt{In our numerical calculations of the velocity residuals, we first construct grids in the disk-plane coordinate $(r,\phi)$ and compute $v_{\rm los,0}$ and $v_{\rm los, \delta}$ for Model 0 and 1, respectively. We then project these values onto the sky plane by transforming $(r,\phi)$ into $(x,y)$ for each model. Next, we prepare a uniform grid in the sky coordinates $(x,y)$ and resample $v_{\rm los}$ onto this grid by interpolating the computed values for the two models, computing their difference as the velocity residuals. Finally, we deproject the velocity residuals back into the disk-plane coordinates $(r,\phi)$ by resampling from $(x,y)$ to $(r,\phi)$ using the geometry of the fiducial model, Model 0. Model 1 geometry may be used at this stage, but the difference is order of $O(\delta ^2)$, as discussed in Section \ref{sec:vel_res_obs}.}

\begin{figure*}
\centering
\includegraphics[width=0.7\linewidth]{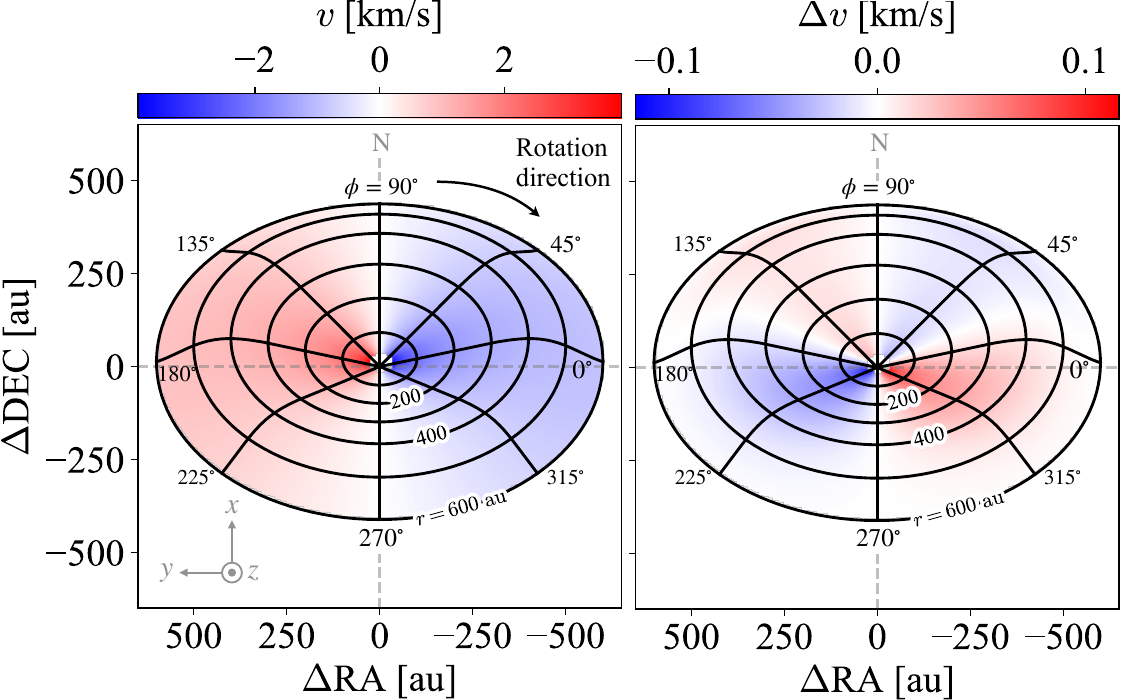}
\caption{Sky-plane line-of-sight velocity (left) and residuals (right) from numerical calculations for a height perturbation of $\delta h(r)=0.1\,h(r)$.   }
\label{fig:comp}
\end{figure*}

\begin{figure*}
\begin{center}
\includegraphics[width=0.9\linewidth]{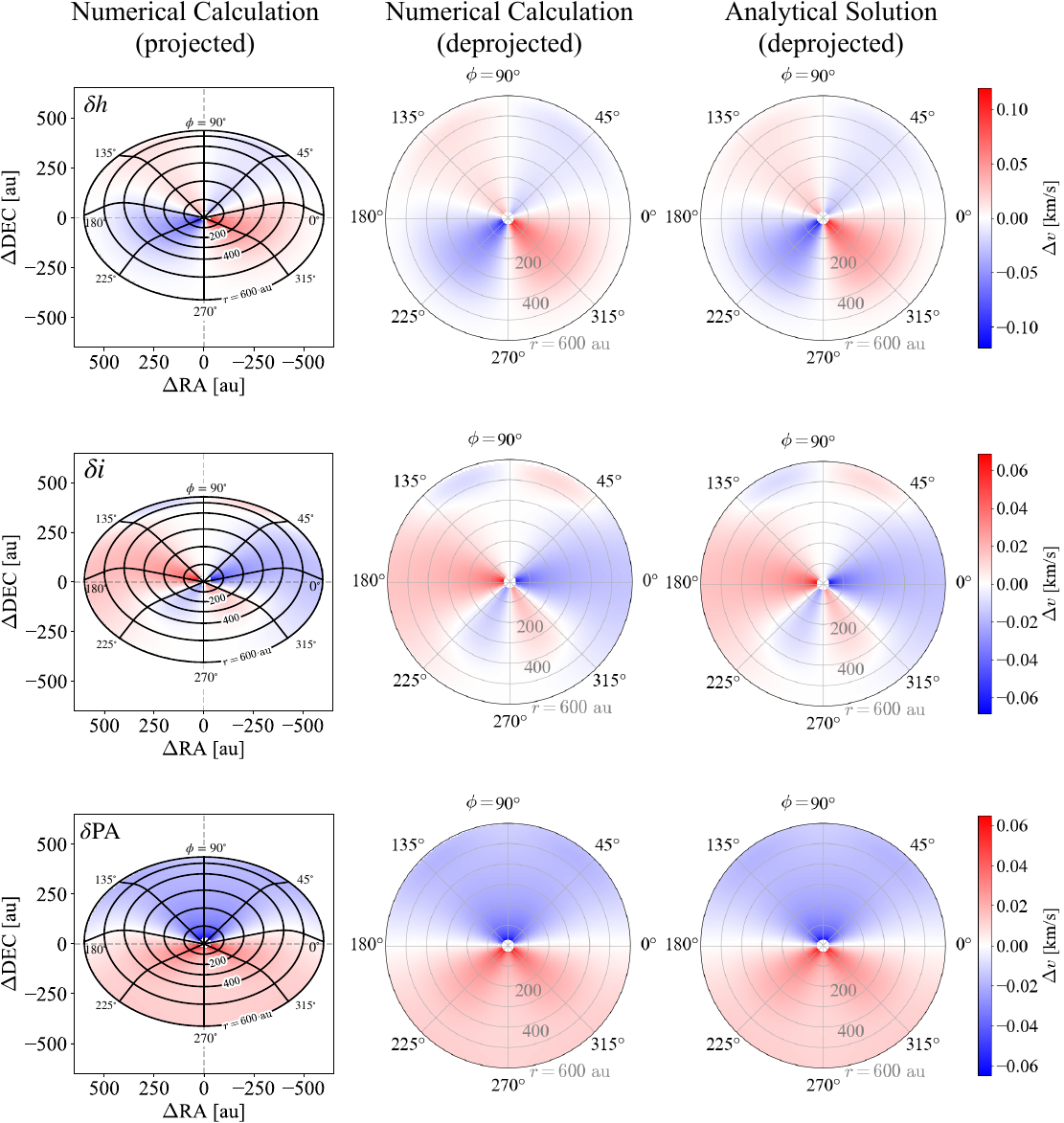}
\caption{Comparison of line-of-sight velocity residuals from numerical calculation (left column) and analytical solution (right column) for three perturbations: $\delta h(r)=0.1\,h(r)$ (top row), $\delta i=1^\circ$ (middle row), and $\delta\mathrm{PA}=1^\circ$ (bottom row).  }
\label{fig:comp_depro}
\end{center}
\end{figure*}

\begin{figure*}
\begin{center}
\includegraphics[width=0.42\linewidth]{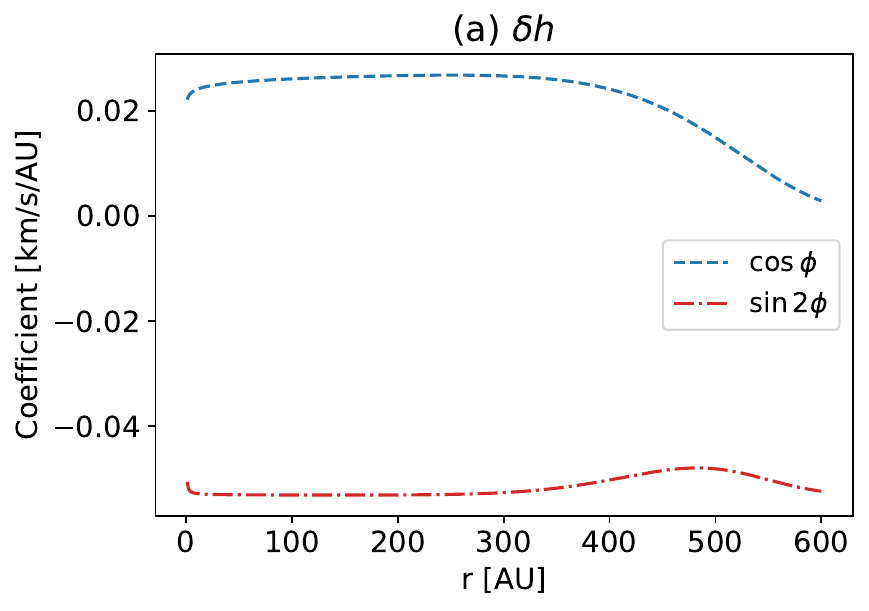}
\includegraphics[width=0.4\linewidth]{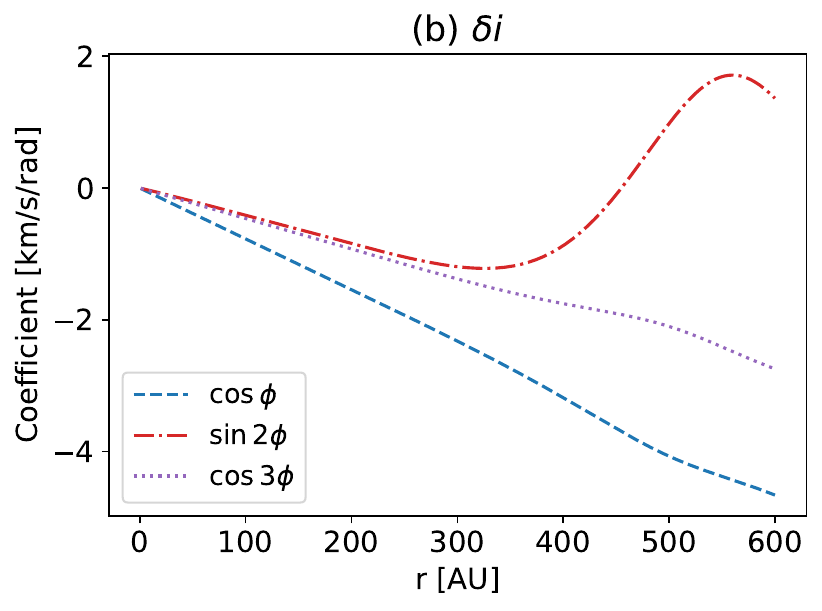}
\includegraphics[width=0.42\linewidth]{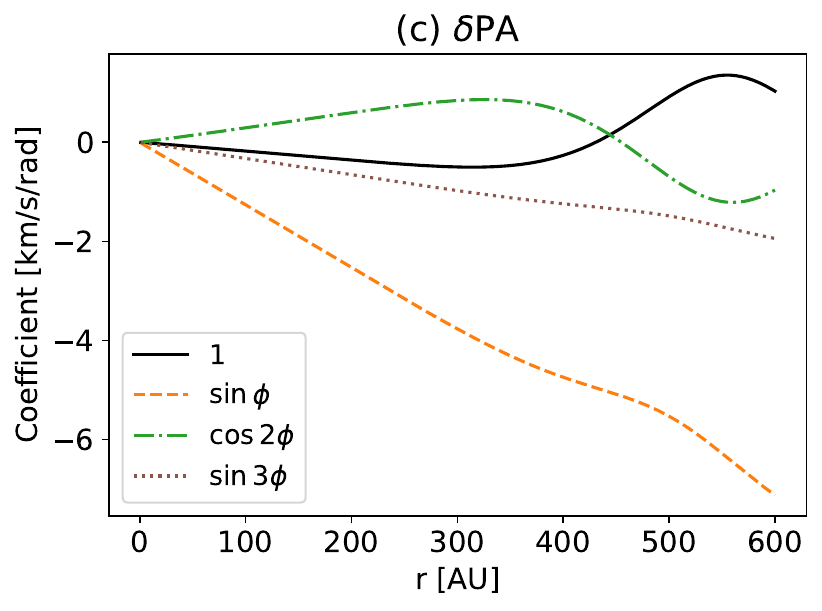}
\includegraphics[width=0.49\linewidth]{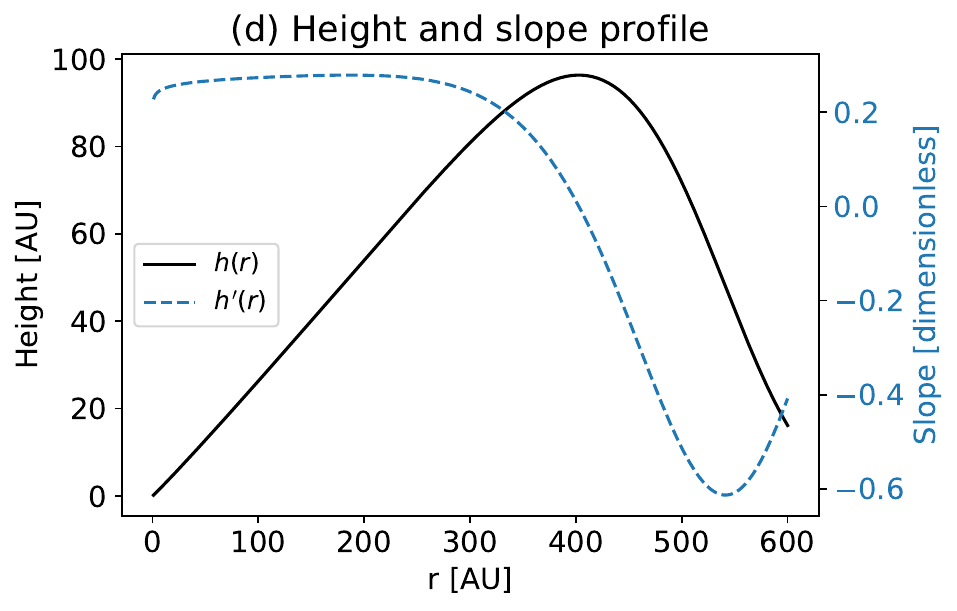}
\caption{ \aizwt{(a-c)} Radial profiles of the key Fourier coefficients in Equation~(\ref{eq:delta_final_clean}) for perturbations in (a) disk height $\delta h(r)$, (b) inclination $\delta i$, and (c) position angle $\delta\mathrm{PA}$. \aizwt{(d) Fiducial disk surface height profile $h_0(r)$ (solid) and its radial derivative $h_0'(r)$ (dashed). }}
\label{fig:coeff}
\end{center}
\end{figure*}

Figures~\ref{fig:comp} shows the numerically computed velocity fields and residuals, defined as the difference between Model 1 (true) and Model 0 (approximate), in the projected frame. The models differ by a perturbation in disk surface height, with $\delta h(r) = 0.1 h(r)$ applied in Model 1. Figure~\ref{fig:comp_depro} compare the numerically computed velocity residuals with the analytical predictions in the deprojected frame for perturbations in $\delta h$, $\delta i$, and $\delta\mathrm{PA}$. The excellent agreement confirms the validity of the analytical model presented here. Figure~\ref{fig:coeff} plots the radial profiles of the harmonic coefficients, alongside $h_{0}(r)$ and $h_{0}'(r)$. \aizwt{The coefficients $B_i(r)$, $C_i(r)$, $D_i(r)$, $E_i(r)$, and $F_i(r)$ depend on $h_{0}(r)$ and $h_{0}'(r)$, as also illustrated in the figure. The strongest modes are $\sin2\phi$, $\cos\phi$, and $\sin\phi$ for $\delta h$, $\delta i$, and $\delta\mathrm{PA}$, respectively, although additional modes can also have non-negligible contributions, as shown in the figures.}

\subsection{Comments on previous models in \cite{winter2025}} \label{sec:comment_winter}

We reexamine the analytic approach of \cite{winter2025}. They begin from the line-of-sight velocity
\begin{equation}
  v_{\rm los}(r,\phi)
  = \sqrt{\frac{GM_\star}{r}}
    \sin i \cos (\phi-\mathrm{PA}),  \label{eq:los_incorrect}
\end{equation}
setting $h(r)=0$ and linearize in $\delta i,\delta\mathrm{PA}$ to obtain
\begin{equation}
\delta v =  \sqrt{\frac{GM_\star}{r}}
      \left ( \cos i \cos (\phi-\mathrm{PA})  \delta i + \sin i\sin (\phi-\mathrm{PA}) \delta \mathrm{PA}\right).  \label{eq:winter}
\end{equation}

However, $\mathrm{PA}$ denotes a rotation of the sky-plane frame and is not part of the disk-plane azimuth $\phi$, so it is, strictly speaking, incorrect to use $\cos (\phi - \mathrm{PA})$. While \cite{winter2025} set $\mathrm{PA}=0$ assuming ``without loss of generality", which reduces the expression to $\cos \phi$ as in our study, the use of $(\phi-\mathrm{PA})$ still introduces an unintended $\partial/\partial\mathrm{PA}$ term in Equation~(\ref{eq:winter}). Although $\delta\mathrm{PA}$ does produce a $\sin\phi$ via mapping errors in our model, this effect is distinct from the derivative of $\cos(\phi-\mathrm{PA})$ in \cite{winter2025}.  

\cite{winter2025} also neglect errors arising from the projection-deprojection mapping. Even when $h(r)=h'(r)=0$ as in \cite{winter2025}, the deprojection generates coefficients $C_0(r), C_2(r), D_1(r)$, and $D_3(r)$ with $A(r)$ being $1/2r$. These neglected contributions produce azimuthal harmonics with $m=1-3$ that are non-negligible, and in particular the $m=1$ components have amplitudes that differ from those adopted by \cite{winter2025}.

These factors can qualitatively alter the interpretation of the observed velocity fields.  Moreover, variations in the disk surface height, not considered in \cite{winter2025}, can introduce significant residuals. Although \citet{winter2025} make the important contribution of introducing a perturbative expansion of velocity residuals in protoplanetary disks, a comprehensive reanalysis of their exoALMA results is warranted to ensure an accurate interpretation of the observational data.

\aizwt{
\section{On inference of perturbed geometry from observations}
\label{sec:determine_perturb}
We describe how perturbations $(\delta h(r), \delta i(r), \delta \text{PA}(r))$ can be recovered from observed velocity residuals. Since the problem reduces to a linear regression model, as discussed in the following subsections, the posterior distribution and its sampling can be derived analytically, assuming Gaussian noise and Gaussian priors, analogous to other problems such as global mapping of exoplanets from reflected light \citep{kawahara2020} and axisymmetric modeling of dust emission of protoplanetary disks \citep{aizawa2024}. 

\subsection{Inference of perturbed geometry at individual annulus}
We rewrite Equation~(\ref{eq:delta_final_clean}) as a Fourier series up to the first order in the perturbation: 
\begin{align}
(-1)^{\varepsilon+1}  \frac{\Delta(r,\phi)\,\delta v_{\rm res,depro}(r,\phi)}{V_{K,0}(r)}
&= a_{0}(r)
+ \sum_{n=1}^{3}\bigl[a_{n}(r) \cos n\phi + b_{n}(r) \sin n\phi\bigr], 
\label{eq:expand_fourier}
\end{align}
where the coefficients are written as 
\begin{align}
\begin{pmatrix}
a_0(r) \\[1pt]
a_1(r) \\[1pt]
a_2(r) \\[1pt]
a_3(r) \\[1pt]
b_1(r) \\[1pt]
b_2(r) \\[1pt]
b_3(r)
\end{pmatrix}
&=
\bm{X}(r) \bm{\theta}_i, \\
\bm{X}(r) &\equiv
\begin{pmatrix}
0                  & 0                    & D_0(r) \\[1pt]
F_0(r)             & C_0(r)+E_0(r)        & 0 \\[1pt]
0                  & 0                    & D_2(r) \\[1pt]
0                  & C_2(r)               & 0      \\[1pt]
0                  & 0                    & D_1(r) \\[1pt]
B_0(r)+F_1(r)      & C_1(r)+E_1(r)        & 0      \\[1pt]
0                  & 0                    & D_3(r)
\end{pmatrix}, \\
\bm{\theta}(r) &= \begin{pmatrix}
\delta h(r) \\[1pt]
\delta i(r) \\[1pt]
\delta\mathrm{PA}(r)
\end{pmatrix}.  \label{eq:linear_design}
\end{align}
Here, $\bm{X}(r)$ corresponds to the design matrix in the linear regression model. 

On the other hand, from the residual velocity field, we can measure the observed Fourier coefficients at radii $r_i$ ($i=1,2,\ldots,N_r$), 
which we denote by 
\begin{equation}
\bm{d} (r_i) \equiv 
    \begin{pmatrix}
a_{\rm obs, 0}(r_i) \\[1pt]
a_{\rm obs, 1}(r_i) \\[1pt]
a_{\rm obs, 2}(r_i) \\[1pt]
a_{\rm obs, 3}(r_i) \\[1pt]
b_{\rm obs, 1}(r_i) \\[1pt]
b_{\rm obs, 2}(r_i) \\[1pt]
b_{\rm obs, 3}(r_i)
\end{pmatrix}.
\end{equation}
Considering measurement noise, we then have
\begin{equation}
\bm{d}_i = \bm{X}_i
\bm{\theta}_i
+ \bm{\epsilon}_i, 
\label{eq:i_d_noise}
\end{equation}
where we denote $\bm{d}_i \equiv \bm{d} (r_i), \bm{X}_i \equiv \bm{X}(r_i), \bm{\theta} = \bm{\theta} (r_i)$. 
Here, $\bm{\epsilon}_i$ denotes the noise vector associated with $\bm{d}_i$, which is assumed to follow a multivariate normal distribution with the 
mean vector of $\bm{0}$ and the covariance matrix of $\bm{\Sigma}_i$: 
\begin{equation}
\bm{\epsilon}_i \sim \mathcal{N}(\bm{0}, \bm{\Sigma}_i). 
\end{equation}

The likelihood for $\bm{\theta}_i$ is therefore
\begin{equation}
p(\bm{d}_i | \bm{\theta}_i)= \mathcal{N}(\bm{d}_i | \bm{X}_i \bm{\theta}_i, \Sigma_i).
\end{equation}
Maximizing this likelihood yields the maximum-likelihood estimate (MLE):
\begin{equation}
\bm{\theta}_{{\rm MLE}, i} = (\bm{X}_i^T \bm{\Sigma}_i^{-1}\bm{X}_i)^{-1} \bm{X}_i^T  \bm{\Sigma}_i^{-1} \bm{d}_i. \label{eq:mle_for_i}
\end{equation}

By imposing a Gaussian prior on $\bm{\theta}_i$ with the mean vector of $\bm{0}$ and the covariance matrix $\bm{
\Sigma}_{\theta, i}$, we have
\begin{equation}
\bm{\theta}_i \sim \mathcal{N}(\bm{0}, \bm{\Sigma}_{\theta, i}). 
\end{equation}
Then, the posterior distribution can be obtained analytically \citep[e.g.,][]{kawahara2020}: 
\begin{equation}
    p( \bm{\theta}_i | \bm{d}_i)= \mathcal{N}(\bm{\theta}_i | (\bm{X}_i^T \bm{\Sigma}_i^{-1}\bm{X}_i +  \bm{\Sigma}_{\theta, i}^{-1})^{-1} \bm{X}_i^T \bm{\Sigma}_i^{-1} \bm{d}_i, (\bm{X}_i^T \bm{\Sigma}_i^{-1}\bm{X}_i + \bm{\Sigma}_{\theta, i}^{-1})^{-1}).
\end{equation}
This gives the full posterior at each radius $r_i$ given $(\bm{d}_i, \bm{\Sigma}_i)$.

\subsection{Joint inference of $(\delta h(r), \delta i(r), \delta\mathrm{PA}(r))$ across annuli }
We now extend the inference to all annuli simultaneously by stacking Equation~(\ref{eq:i_d_noise}) across radii $r_1, \ldots, r_{N_r}$:
\begin{align}
\bm{d}
&=\operatorname{diag} \big(\bm{X}_1, \bm{X}_2, \ldots, \bm{X}_{N_r} \big)
\begin{pmatrix}
\bm{\theta}_1 \\ 
\bm{\theta}_2 \\ 
\vdots \\
\bm{\theta}_{N_r}
\end{pmatrix}
+ \bm{\epsilon},  \label{eq:model_for_joint}
\end{align}
where 
\begin{align}
\bm{d}\equiv 
\begin{pmatrix}
\bm{d}_1 \\
\bm{d}_2 \\
\vdots \\
\bm{d}_{N_r}
\end{pmatrix}, 
\quad 
\operatorname{diag} \big(\bm{X}_1, \bm{X}_2, \ldots, \bm{X}_{N_r} \big) = 
\begin{pmatrix}
\bm{X}_1 & & & \\
& \bm{X}_2 & & \\
& & \ddots & \\
& & & \bm{X}_{N_r}
\end{pmatrix}, \quad 
\bm{\epsilon}\equiv 
\begin{pmatrix}
\bm{\epsilon}_1 \\
\bm{\epsilon}_2 \\
\vdots \\
\bm{\epsilon}_{N_r}
\end{pmatrix}.
\end{align}

We define a permutation matrix $\bm{P}$ such that the global parameter vector 
\begin{equation}
\bm{\theta} 
\equiv 
\bm{P}
\begin{pmatrix}
\bm{\theta}_1 \\ 
\bm{\theta}_2 \\ 
\vdots \\
\bm{\theta}_{N_r}
\end{pmatrix}
= 
\begin{pmatrix}
\delta h(r_1) \\ \vdots \\ \delta h(r_{N_r}) \\
\delta i(r_1) \\ \vdots \\ \delta i(r_{N_r}) \\
\delta \mathrm{PA}(r_1) \\ \vdots \\ \delta \mathrm{PA}(r_{N_r})
\end{pmatrix}.
\end{equation}
Equation~(\ref{eq:model_for_joint}) can then be written compactly as
\begin{align}
\bm{d}= \bm{Y} \bm{\theta} + \bm{\epsilon},
\end{align}
where
\begin{equation}
  \bm{Y} \equiv \operatorname{diag} \big(\bm{X}_1, \bm{X}_2, \ldots, \bm{X}_{N_r} \big)  \bm{P^{-1}}.
\end{equation}

Assuming independent Gaussian noise across annuli, we obtain 
\begin{equation}
\bm{\epsilon} \sim \mathcal{N}\left(
\bm{0},\bm{\Sigma}_{d}
\right) \equiv \mathcal{N} \left(
\bm{0} , 
\operatorname{diag} \big(\bm{\Sigma}_1, \bm{\Sigma}_2, \ldots, \bm{\Sigma}_{N_r}\big) \right). 
\end{equation}
The MLE of $\bm{\theta}$ is then 
\begin{equation}\bm{\theta}_{\rm MLE} = (\bm{Y}^T \bm{\Sigma}^{-1}_d\bm{Y})^{-1} \bm{Y}^T  \bm{\Sigma}^{-1}_d \bm{d}.
\end{equation}
Because the noise model assumes independence across annuli, this expression is equivalent to solving Equation~(\ref{eq:mle_for_i}) independently at each radius, differing only by the ordering imposed through the permutation matrix. 

For the joint inference across annuli, we impose a Gaussian prior on $\bm{\theta}$ 
that is separable among the perturbed geometries: 
\begin{align}
p(\bm{\theta}) = \mathcal{N}\left(\bm{\theta} \middle|
\bm{0}, \operatorname{diag} \big(\bm{\Sigma}_h, \bm{\Sigma}_i, \bm{\Sigma}_{\rm PA} \big)
\right), 
\end{align}
where $\bm{\Sigma}_h, \bm{\Sigma}_i, \bm{\Sigma}_{\rm PA}$ are covariance matrices for $\delta h$, $\delta i$, and $\delta {\rm PA}$, respectively. 

When these prior covariances are defined through kernel functions parameterized by hypyerparameters $\bm{\phi}$, we write
\begin{equation}
\bm{\Sigma} (\bm{\phi}) = \operatorname{diag} \big(\bm{\Sigma}_h, \bm{\Sigma}_i, \bm{\Sigma}_{\rm PA} \big). 
\end{equation}
For example, adopting a radius basis function kernel for each introduces two hyperparameters per geometry; there are six parameters in total. 

The posterior distribution of $\bm{\theta}$ is again given in closed form:
\begin{equation}
    p(\bm{\theta} | \bm{d}) = \mathcal{N}(\bm{\theta} | (\bm{Y}^T \bm{\Sigma}_{d}^{-1}\bm{Y} +  \bm{\Sigma}(\bm{\phi})^{-1} )^{-1} \bm{Y}^T \bm{\Sigma}_{d}^{-1} \bm{d}, (\bm{Y}^T \bm{\Sigma}_{d}^{-1}\bm{Y} +  \bm{\Sigma}(\bm{\phi})^{-1} )^{-1}). \label{eq:post_theta_d}
\end{equation}
and the marginal likelihood for the hyperparameters $\bm{\phi}$, obtained by integrating out $\bm{\theta}$, can be computed analytically \citep[e.g.,][]{kawahara2020}: 
\begin{equation}
    p(\bm{d}  |\bm{\phi})= p(\bm{d} | \bm{0}, \bm{\Sigma}_d +\bm{Y} \bm{\Sigma}(\bm{\phi})\bm{Y}^T).
\end{equation}
We can obtain samples from the joint posterior $p(\bm{\theta},\bm{\phi} | \bm{d})$ via blocked Gibbs sampling. 
Assuming a prior on $p(\bm{\phi})$, we first draw samples $\bm{\phi}^{\dagger} \sim p(\bm{\phi} | \bm{d} ) \propto  p(\bm{d}  |\bm{\phi}) p(\bm{\phi}) $,  and subsequently take samples from $\bm{\theta}^{\dagger} \sim p(\bm{\theta}| \bm{d}, \bm{\phi}^{\dagger})$ using Equation (\ref{eq:post_theta_d}). Iterating these steps yields samples from the full posterior distribution $p(\bm{\theta}, \bm{\phi} | \bm{d})$. 
}

\section{Conclusion and Discussion} \label{sec:conclusions}

We present a first-order analytical model for line-of-sight velocity residuals in flared, nearly axisymmetric disks subject to small perturbations $(\delta h(r),\delta i(r),\delta\mathrm{PA}(r))$.  By accounting for both projection and deprojection, we show that the normalized residual field, derivable from observations and the fiducial model, decomposes into azimuthal harmonics up to $m=3$.  We also present the framework for recovering the perturbation profiles $\delta h(r)$, $\delta i(r)$ and $\delta\mathrm{PA}(r)$ from the data. The warped-disk model employed by \cite{winter2025} is susceptible to inaccuracies arising from simplified assumptions, and the reanalysis of the exoALMA data within our framework would thus enable more reliable interpretation.

In this work, we adopt an idealized, noise-free case without beam smearing and assume that emission originates exclusively from the upper disk surface; however, neither assumption holds in real observations. We also neglect radial pressure-gradient corrections or self-gravity terms \citep[e.g.,][]{pinte2018,longarini2025,stadler2025}, which alter the azimuthal velocity profile, as well as any radial or vertical flow components \citep{rosenfeld2014,teague2019,zuleta2024}. 

Several extensions to our model are possible. First, allowing the fiducial inclination and position angle to vary with radius, $i_0(r)$ and $\mathrm{PA}_0(r)$, and iteratively updating  $\bigl[h_0(r), i_0(r), \mathrm{PA}_0(r)\bigr] \to  \bigl[h_0(r)+\delta h(r), i_0(r)+\delta i(r), \mathrm{PA}_0(r)+\delta\mathrm{PA}(r)\bigr]$ would enable convergence to the true geometry even in strongly warped disks.  This requires adding the corresponding partial derivatives of $i(r)$ and $\mathrm{PA}(r)$ in the expressions for $f_r$ and $g_r$ in Equations~(\ref{eq:f_der}) and (\ref{eq:g_der}) and all subsequent terms.  Second, our perturbative framework can be applied to other observables, such as line intensities or widths, as functions of $(r,\phi)$ by replacing the partial derivatives in Equations~(\ref{eq:model_dif_0}) and (\ref{eq:project_deproject_eq}) accordingly. The velocity fields can be also modified to incorporate pressure-gradient corrections, self-gravity terms, and vertical motions. Such extensions would enable a more robust inference of disk parameters and the identification of non-axisymmetric features (e.g.\ spirals). We leave them to future work.

\section*{Acknowledgements}
We acknowledge fruitful discussions with Munetake Momose. \aizwt{We thank the referee for the constructive comments that greatly improved the manuscript.} This work was supported by JSPS KAKENHI grant number 25K17431. We acknowledge the use of ChatGPT (GPT-5; OpenAI) for grammatical and clarity improvements throughout the manuscript.

\appendix 
\section{Explicit expressions for derivatives and coefficients} 
\subsection{Derivatives of $f$ and $g$ with respect to $r$, $\phi$, $h$, $i$, and $\mathrm{PA}$} \label{sec:3_detailed_dr_dphi}

Each term in Equation (\ref{eq:partial_deri_eq}) with respect to $r,\phi$ derivatives is given by: 
\begin{align}
f_r &= (\cos\phi\cos\mathrm{PA}_0-\cos i_0\sin\phi\sin\mathrm{PA}_0) - h_0'(r)\sin i_0\sin\mathrm{PA}_0,\quad 
f_\phi = - r(\sin\phi\cos\mathrm{PA}_0+\cos i_0\cos\phi\sin\mathrm{PA}_0), \label{eq:f_der}\\
g_r &= (\cos\phi\sin\mathrm{PA}_0+\cos i_0\sin\phi\cos\mathrm{PA}_0)+h_0'(r)\sin i_0\cos\mathrm{PA}_0,\quad 
g_\phi = r(- \sin\phi\sin\mathrm{PA}_0+\cos i_0\cos\phi\cos\mathrm{PA}_0)   \label{eq:g_der}
\end{align}

The each of term in Equation (\ref{eq:partial_deri_eq}) with respect to $h, i, \mathrm{PA}$ derivatives is given by: 
\begin{align}
&f_h=-\sin i_0\sin\mathrm{PA}_0, \quad
f_{\mathrm{PA}}=-r(\cos\phi\sin\mathrm{PA}_0+\cos i_0\sin\phi\cos\mathrm{PA}_0)
- h_0(r)\sin i_0\cos\mathrm{PA}_0,\\
&f_{i}=r\sin i_0\sin\phi\sin\mathrm{PA}_0
- h_0(r)\cos i_0\sin\mathrm{PA}_0,\\
&g_h= \sin i_0\cos\mathrm{PA}_0, \quad
g_{\mathrm{PA}}=r(\cos\phi\cos\mathrm{PA}_0-\cos i_0\sin\phi\sin\mathrm{PA}_0)
- h_0(r)\sin i_0\sin\mathrm{PA}_0,\\
&g_i =-r\sin i_0\sin\phi\cos\mathrm{PA}_0
+ h_0(r)\cos i_0\cos\mathrm{PA}_0.
\end{align}

In addition, we find: 
\begin{align}
r_h \Delta  &= -r \sin i_0 \sin \phi, \quad  
\phi_h \Delta  =- \sin i_0 \cos \phi, \quad 
r_i \Delta = -r \sin \phi( h_0(r) \cos i_0- r \sin i_0 \sin\phi) \\
\phi_i \Delta &= r \sin i_0 \cos \phi \sin \phi - h_0(r) \cos i_0\cos \phi, \quad 
r_{{\mathrm{PA}_0}}  \Delta = -r \cos \phi \sin i_0 (r \sin \phi \sin i_0 - h \cos i_0) \\
\phi_{{\mathrm{PA}_0}} \Delta &= -h_0(r) \sin i_0 ( h_0'(r) \sin i_0 + \cos i_0 \sin \phi) - r h_0'(r) \sin i_0 \cos i_0\sin \phi - r ( \cos^2 \phi + \cos^2 i_0 \sin ^2 \phi) 
\end{align}

\subsection{Explicit forms of $B_i(r)$-$F_i(r)$ }\label{sec:3_depro_error}

The explicit functional forms of the coefficients for $B_i(r)$-$F_i(r)$ are 
\begin{align}
B_0(r) &= -\left[\frac{\sin i_0 (rA(r)+1)}{2}\right], \quad 
C_0(r)=  \left[\frac{r+r^2A(r)}{4}\right] \sin i_0, \quad 
C_1(r) = - \left[ \frac{1+rA(r)}{2}\right] h_0(r)\cos i_0\\
C_2(r)&=- \left[\frac{(1+rA(r))}{4} \right]r  \sin i_0, \quad 
D_0(r) =  \frac{\sin i_0\cos i}{2}\bigl[r A(r) h_0(r) - h_0(r) - rh_0'(r)\bigr], \quad  \\
D_1(r) &= - h_0(r)h_0'(r)\sin^2 i_0
         - \frac{r^2A(r)\sin^2 i_0 + r(1+3\cos^2 i_0)}{4}, \quad
D_2(r) =  \frac{\sin i_0\cos i}{2}\bigl[r A(r) h_0(r)+ h_0(r) + rh_0'(r)\bigr], \quad \\
D_3(r) &= -\frac{r\bigl(rA(r)+1\bigr)\sin^2 i_0}{4},  \quad E_0(r) \equiv - r\cos i_0\cot i, \quad \quad E_1(r) \equiv  - \frac{r h_0'(r)\cos i_0}{2}, \quad \\
F_0(r) &\equiv  \left[\frac{3h_0(r)  r \cos i_0}{2(r^2 + h_0^2(r))}\right], \quad F_1(r) \equiv \left[\frac{3h_0(r) h_0'(r)\sin i_0 }{\aizw{4}(r^2 + h_0^2(r))}\right]. 
\end{align}

\subsection{Fourier decomposition of the velocity residual}
\label{sec:3_fourier_exp}
The Fourier expansion for Equation (\ref{eq:delta_final_clean}) is  
\begin{align}
(-1)^{\varepsilon+1} \frac{\delta v_{\rm res,depro}(r,\phi)\,\Delta(r,\phi)}{V_{K,0}(r)}
&=
G_0(r)
+ G_1(r)\cos\phi
+ G_2(r)\cos2\phi
+ G_3(r)\cos3\phi \\
&+ H_1(r)\sin\phi
+ H_2(r)\sin2\phi
+ H_3(r)\sin3\phi,
\end{align}
where
\[
\begin{aligned}
G_0(r)&=D_{0}(r)\,\delta\mathrm{PA},\\
G_1(r)&=F_{0}(r)\,\delta h
        +\bigl[C_{0}(r)+E_{0}(r)\bigr]\,\delta i ,\\
G_2(r)&=D_{2}(r)\,\delta\mathrm{PA},\\
G_3(r)&=C_{2}(r)\,\delta i,\\
H_1(r)&=D_{1}(r)\,\delta\mathrm{PA},\\
H_2(r)&=\bigl[B_{0}(r)+F_{1}(r)\bigr]\,\delta h
        +\bigl[C_{1}(r)+E_{1}(r)\bigr]\,\delta i,\\
H_3(r)&=D_{3}(r)\,\delta\mathrm{PA}.
\end{aligned}
\]




\bibliographystyle{aasjournal}
\bibliography{ref}

@ARTICLE{izquierdo2021,
       author = {{Izquierdo}, A.~F. and {Testi}, L. and {Facchini}, S. and {Rosotti}, G.~P. and {van Dishoeck}, E.~F.},
        title = "{The Disc Miner. I. A statistical framework to detect and quantify kinematical perturbations driven by young planets in discs}",
      journal = {\aap},
         year = 2021,
        month = jun,
       volume = {650},
          eid = {A179},
        pages = {A179},
          doi = {10.1051/0004-6361/202140779},
archivePrefix = {arXiv},
       eprint = {2104.09596},
 primaryClass = {astro-ph.EP},
       adsurl = {https://ui.adsabs.harvard.edu/abs/2021A&A...650A.179I}
}

@ARTICLE{pinte2018,
       author = {{Pinte}, C. and {Price}, D.~J. and {M{\'e}nard}, F. and {Duch{\^e}ne}, G. and {Dent}, W.~R.~F. and {Hill}, T. and {de Gregorio-Monsalvo}, I. and {Hales}, A. and {Mentiplay}, D.},
        title = "{Kinematic Evidence for an Embedded Protoplanet in a Circumstellar Disk}",
      journal = {\apjl},
         year = 2018,
        month = jun,
       volume = {860},
       number = {1},
          eid = {L13},
        pages = {L13},
          doi = {10.3847/2041-8213/aac6dc},
archivePrefix = {arXiv},
       eprint = {1805.10293},
 primaryClass = {astro-ph.SR},
       adsurl = {https://ui.adsabs.harvard.edu/abs/2018ApJ...860L..13P}
}

@ARTICLE{teague2019,
       author = {{Teague}, Richard and {Bae}, Jaehan and {Bergin}, Edwin A.},
        title = "{Meridional flows in the disk around a young star}",
      journal = {\nat},
         year = 2019,
        month = oct,
       volume = {574},
       number = {7778},
        pages = {378-381},
          doi = {10.1038/s41586-019-1642-0},
archivePrefix = {arXiv},
       eprint = {1910.06980},
 primaryClass = {astro-ph.EP},
       adsurl = {https://ui.adsabs.harvard.edu/abs/2019Natur.574..378T}
}

@ARTICLE{law2022,
       author = {{Law}, Charles J. and {Crystian}, Sage and {Teague}, Richard and {{\"O}berg}, Karin I. and {Rich}, Evan A. and {Andrews}, Sean M. and {Bae}, Jaehan and {Flaherty}, Kevin and {Guzm{\'a}n}, Viviana V. and {Huang}, Jane and {Ilee}, John D. and {Kastner}, Joel H. and {Loomis}, Ryan A. and {Long}, Feng and {P{\'e}rez}, Laura M. and {P{\'e}rez}, Sebasti{\'a}n and {Qi}, Chunhua and {Rosotti}, Giovanni P. and {Ru{\'\i}z-Rodr{\'\i}guez}, Dary and {Tsukagoshi}, Takashi and {Wilner}, David J.},
        title = "{CO Line Emission Surfaces and Vertical Structure in Midinclination Protoplanetary Disks}",
      journal = {\apj},
         year = 2022,
        month = jun,
       volume = {932},
       number = {2},
          eid = {114},
        pages = {114},
          doi = {10.3847/1538-4357/ac6c02},
archivePrefix = {arXiv},
       eprint = {2205.01776},
 primaryClass = {astro-ph.EP},
       adsurl = {https://ui.adsabs.harvard.edu/abs/2022ApJ...932..114L}
}

@ARTICLE{panequecarreno2023,
       author = {{Stapper}, L.~M. and {Hogerheijde}, M.~R. and {van Dishoeck}, E.~F. and {Paneque-Carre{\~n}o}, T.},
        title = "{A dichotomy in group II Herbig disks. ALMA gas disk height measurements show both shadowed large vertically extended disks and compact flat disks}",
      journal = {\aap},
         year = 2023,
        month = jan,
       volume = {669},
          eid = {A158},
        pages = {A158},
          doi = {10.1051/0004-6361/202245137},
archivePrefix = {arXiv},
       eprint = {2211.12532},
 primaryClass = {astro-ph.EP},
       adsurl = {https://ui.adsabs.harvard.edu/abs/2023A&A...669A.158S}
}

@ARTICLE{oberg2021,
       author = {{{\"O}berg}, Karin I. and {Guzm{\'a}n}, Viviana V. and {Walsh}, Catherine and {Aikawa}, Yuri and {Bergin}, Edwin A. and {Law}, Charles J. and {Loomis}, Ryan A. and {Alarc{\'o}n}, Felipe and {Andrews}, Sean M. and {Bae}, Jaehan and {Bergner}, Jennifer B. and {Boehler}, Yann and {Booth}, Alice S. and {Bosman}, Arthur D. and {Calahan}, Jenny K. and {Cataldi}, Gianni and {Cleeves}, L. Ilsedore and {Czekala}, Ian and {Furuya}, Kenji and {Huang}, Jane and {Ilee}, John D. and {Kurtovic}, Nicolas T. and {Le Gal}, Romane and {Liu}, Yao and {Long}, Feng and {M{\'e}nard}, Fran{\c{c}}ois and {Nomura}, Hideko and {P{\'e}rez}, Laura M. and {Qi}, Chunhua and {Schwarz}, Kamber R. and {Sierra}, Anibal and {Teague}, Richard and {Tsukagoshi}, Takashi and {Yamato}, Yoshihide and {van't Hoff}, Merel L.~R. and {Waggoner}, Abygail R. and {Wilner}, David J. and {Zhang}, Ke},
        title = "{Molecules with ALMA at Planet-forming Scales (MAPS). I. Program Overview and Highlights}",
      journal = {\apjs},
         year = 2021,
        month = nov,
       volume = {257},
       number = {1},
          eid = {1},
        pages = {1},
          doi = {10.3847/1538-4365/ac1432},
archivePrefix = {arXiv},
       eprint = {2109.06268},
 primaryClass = {astro-ph.EP},
       adsurl = {https://ui.adsabs.harvard.edu/abs/2021ApJS..257....1O}
}

@ARTICLE{pinte2020,
       author = {{Pinte}, C. and {Price}, D.~J. and {M{\'e}nard}, F. and {Duch{\^e}ne}, G. and {Christiaens}, V. and {Andrews}, S.~M. and {Huang}, J. and {Hill}, T. and {van der Plas}, G. and {Perez}, L.~M. and {Isella}, A. and {Boehler}, Y. and {Dent}, W.~R.~F. and {Mentiplay}, D. and {Loomis}, R.~A.},
        title = "{Nine Localized Deviations from Keplerian Rotation in the DSHARP Circumstellar Disks: Kinematic Evidence for Protoplanets Carving the Gaps}",
      journal = {\apjl},
         year = 2020,
        month = feb,
       volume = {890},
       number = {1},
          eid = {L9},
        pages = {L9},
          doi = {10.3847/2041-8213/ab6dda},
archivePrefix = {arXiv},
       eprint = {2001.07720},
 primaryClass = {astro-ph.SR},
       adsurl = {https://ui.adsabs.harvard.edu/abs/2020ApJ...890L...9P}
}

@ARTICLE{law2021,
       author = {{Law}, Charles J. and {Teague}, Richard and {Loomis}, Ryan A. and {Bae}, Jaehan and {{\"O}berg}, Karin I. and {Czekala}, Ian and {Andrews}, Sean M. and {Aikawa}, Yuri and {Alarc{\'o}n}, Felipe and {Bergin}, Edwin A. and {Bergner}, Jennifer B. and {Booth}, Alice S. and {Bosman}, Arthur D. and {Calahan}, Jenny K. and {Cataldi}, Gianni and {Cleeves}, L. Ilsedore and {Furuya}, Kenji and {Guzm{\'a}n}, Viviana V. and {Huang}, Jane and {Ilee}, John D. and {Le Gal}, Romane and {Liu}, Yao and {Long}, Feng and {M{\'e}nard}, Fran{\c{c}}ois and {Nomura}, Hideko and {P{\'e}rez}, Laura M. and {Qi}, Chunhua and {Schwarz}, Kamber R. and {Soto}, Daniela and {Tsukagoshi}, Takashi and {Yamato}, Yoshihide and {van't Hoff}, Merel L.~R. and {Walsh}, Catherine and {Wilner}, David J. and {Zhang}, Ke},
        title = "{Molecules with ALMA at Planet-forming Scales (MAPS). IV. Emission Surfaces and Vertical Distribution of Molecules}",
      journal = {\apjs},
         year = 2021,
        month = nov,
       volume = {257},
       number = {1},
          eid = {4},
        pages = {4},
          doi = {10.3847/1538-4365/ac1439},
archivePrefix = {arXiv},
       eprint = {2109.06217},
 primaryClass = {astro-ph.GA},
       adsurl = {https://ui.adsabs.harvard.edu/abs/2021ApJS..257....4L}
}

@ARTICLE{casassus2019,
       author = {{Casassus}, Simon and {P{\'e}rez}, Sebasti{\'a}n},
        title = "{Kinematic Detections of Protoplanets: A Doppler Flip in the Disk of HD 100546}",
      journal = {\apjl},
         year = 2019,
        month = oct,
       volume = {883},
       number = {2},
          eid = {L41},
        pages = {L41},
          doi = {10.3847/2041-8213/ab4425},
archivePrefix = {arXiv},
       eprint = {1906.06302},
 primaryClass = {astro-ph.EP},
       adsurl = {https://ui.adsabs.harvard.edu/abs/2019ApJ...883L..41C}
}

@ARTICLE{aizawa2024,
       author = {{Aizawa}, Masataka and {Muto}, Takayuki and {Momose}, Munetake},
        title = "{Revealing asymmetry on mid-plane of protoplanetary disc through modelling of axisymmetric emission: methodology}",
      journal = {\mnras},
         year = 2024,
        month = aug,
       volume = {532},
       number = {2},
        pages = {1361-1390},
          doi = {10.1093/mnras/stae1549},
archivePrefix = {arXiv},
       eprint = {2406.14006},
 primaryClass = {astro-ph.EP},
       adsurl = {https://ui.adsabs.harvard.edu/abs/2024MNRAS.532.1361A}
}

@ARTICLE{casassus2015,
       author = {{Casassus}, S. and {Marino}, S. and {P{\'e}rez}, S. and {Roman}, P. and {Dunhill}, A. and {Armitage}, P.~J. and {Cuadra}, J. and {Wootten}, A. and {van der Plas}, G. and {Cieza}, L. and {Moral}, Victor and {Christiaens}, V. and {Montesinos}, Mat{\'\i}as},
        title = "{Accretion Kinematics through the Warped Transition Disk in HD142527 from Resolved CO(6-5) Observations}",
      journal = {\apj},
         year = 2015,
        month = oct,
       volume = {811},
       number = {2},
          eid = {92},
        pages = {92},
          doi = {10.1088/0004-637X/811/2/92},
archivePrefix = {arXiv},
       eprint = {1505.07732},
 primaryClass = {astro-ph.SR},
       adsurl = {https://ui.adsabs.harvard.edu/abs/2015ApJ...811...92C}
}

@ARTICLE{izquierdo2025,
       author = {{Izquierdo}, Andr{\'e}s F. and {Stadler}, Jochen and {Galloway-Sprietsma}, Maria and {Benisty}, Myriam and {Pinte}, Christophe and {Bae}, Jaehan and {Teague}, Richard and {Facchini}, Stefano and {W{\"o}lfer}, Lisa and {Longarini}, Cristiano and {Curone}, Pietro and {Andrews}, Sean M. and {Barraza-Alfaro}, Marcelo and {Cataldi}, Gianni and {Cuello}, Nicol{\'a}s and {Czekala}, Ian and {Fasano}, Daniele and {Flock}, Mario and {Fukagawa}, Misato and {Garg}, Himanshi and {Hall}, Cassandra and {Hammond}, Iain and {Hilder}, Thomas and {Huang}, Jane and {Ilee}, John D. and {Isella}, Andrea and {Kanagawa}, Kazuhiro and {Lesur}, Geoffroy and {Lodato}, Giuseppe and {Loomis}, Ryan A. and {Orihara}, Ryuta and {Price}, Daniel J. and {Rosotti}, Giovanni and {Testi}, Leonardo and {Yen}, Hsi-Wei and {Wafflard-Fernandez}, Gaylor and {Wilner}, David J. and {Winter}, Andrew J. and {Yoshida}, Tomohiro C. and {Zawadzki}, Brianna},
        title = "{exoALMA. III. Line-intensity Modeling and System Property Extraction from Protoplanetary Disks}",
      journal = {\apjl},
         year = 2025,
        month = may,
       volume = {984},
       number = {1},
          eid = {L8},
        pages = {L8},
          doi = {10.3847/2041-8213/adc439},
archivePrefix = {arXiv},
       eprint = {2504.19986},
 primaryClass = {astro-ph.EP},
       adsurl = {https://ui.adsabs.harvard.edu/abs/2025ApJ...984L...8I}
}

@ARTICLE{rosenfeld2013,
       author = {{Rosenfeld}, Katherine A. and {Andrews}, Sean M. and {Hughes}, A. Meredith and {Wilner}, David J. and {Qi}, Chunhua},
        title = "{A Spatially Resolved Vertical Temperature Gradient in the HD 163296 Disk}",
      journal = {\apj},
         year = 2013,
        month = sep,
       volume = {774},
       number = {1},
          eid = {16},
        pages = {16},
          doi = {10.1088/0004-637X/774/1/16},
archivePrefix = {arXiv},
       eprint = {1306.6475},
 primaryClass = {astro-ph.SR},
       adsurl = {https://ui.adsabs.harvard.edu/abs/2013ApJ...774...16R}
}

@ARTICLE{casassus2022,
       author = {{Casassus}, Simon and {C{\'a}rcamo}, Miguel and {Hales}, Antonio and {Weber}, Philipp and {Dent}, Bill},
        title = "{The Doppler Flip in HD 100546 as a Disk Eruption: The Elephant in the Room of Kinematic Protoplanet Searches}",
      journal = {\apjl},
         year = 2022,
        month = jul,
       volume = {933},
       number = {1},
          eid = {L4},
        pages = {L4},
          doi = {10.3847/2041-8213/ac75e8},
archivePrefix = {arXiv},
       eprint = {2206.03236},
 primaryClass = {astro-ph.EP},
       adsurl = {https://ui.adsabs.harvard.edu/abs/2022ApJ...933L...4C}
}

@ARTICLE{pinte2018_2,
       author = {{Pinte}, C. and {M{\'e}nard}, F. and {Duch{\^e}ne}, G. and {Hill}, T. and {Dent}, W.~R.~F. and {Woitke}, P. and {Maret}, S. and {van der Plas}, G. and {Hales}, A. and {Kamp}, I. and {Thi}, W.~F. and {de Gregorio-Monsalvo}, I. and {Rab}, C. and {Quanz}, S.~P. and {Avenhaus}, H. and {Carmona}, A. and {Casassus}, S.},
        title = "{Direct mapping of the temperature and velocity gradients in discs. Imaging the vertical CO snow line around IM Lupi}",
      journal = {\aap},
         year = 2018,
        month = jan,
       volume = {609},
          eid = {A47},
        pages = {A47},
          doi = {10.1051/0004-6361/201731377},
archivePrefix = {arXiv},
       eprint = {1710.06450},
 primaryClass = {astro-ph.SR},
       adsurl = {https://ui.adsabs.harvard.edu/abs/2018A&A...609A..47P}
}

@ARTICLE{benisty2017,
       author = {{Benisty}, M. and {Stolker}, T. and {Pohl}, A. and {de Boer}, J. and {Lesur}, G. and {Dominik}, C. and {Dullemond}, C.~P. and {Langlois}, M. and {Min}, M. and {Wagner}, K. and {Henning}, T. and {Juhasz}, A. and {Pinilla}, P. and {Facchini}, S. and {Apai}, D. and {van Boekel}, R. and {Garufi}, A. and {Ginski}, C. and {M{\'e}nard}, F. and {Pinte}, C. and {Quanz}, S.~P. and {Zurlo}, A. and {Boccaletti}, A. and {Bonnefoy}, M. and {Beuzit}, J.~L. and {Chauvin}, G. and {Cudel}, M. and {Desidera}, S. and {Feldt}, M. and {Fontanive}, C. and {Gratton}, R. and {Kasper}, M. and {Lagrange}, A. -M. and {LeCoroller}, H. and {Mouillet}, D. and {Mesa}, D. and {Sissa}, E. and {Vigan}, A. and {Antichi}, J. and {Buey}, T. and {Fusco}, T. and {Gisler}, D. and {Llored}, M. and {Magnard}, Y. and {Moeller-Nilsson}, O. and {Pragt}, J. and {Roelfsema}, R. and {Sauvage}, J. -F. and {Wildi}, F.},
        title = "{Shadows and spirals in the protoplanetary disk HD 100453}",
      journal = {\aap},
         year = 2017,
        month = jan,
       volume = {597},
          eid = {A42},
        pages = {A42},
          doi = {10.1051/0004-6361/201629798},
archivePrefix = {arXiv},
       eprint = {1610.10089},
 primaryClass = {astro-ph.EP},
       adsurl = {https://ui.adsabs.harvard.edu/abs/2017A&A...597A..42B}
}

@ARTICLE{marino2015,
       author = {{Marino}, S. and {Perez}, S. and {Casassus}, S.},
        title = "{Shadows Cast by a Warp in the HD 142527 Protoplanetary Disk}",
      journal = {\apjl},
         year = 2015,
        month = jan,
       volume = {798},
       number = {2},
          eid = {L44},
        pages = {L44},
          doi = {10.1088/2041-8205/798/2/L44},
archivePrefix = {arXiv},
       eprint = {1412.4632},
 primaryClass = {astro-ph.EP},
       adsurl = {https://ui.adsabs.harvard.edu/abs/2015ApJ...798L..44M}
}

@ARTICLE{zuleta2024,
       author = {{Zuleta}, A. and {Birnstiel}, T. and {Teague}, R.},
        title = "{Kinematical signatures: Distinguishing between warps and radial flows}",
      journal = {\aap},
         year = 2024,
        month = dec,
       volume = {692},
          eid = {A56},
        pages = {A56},
          doi = {10.1051/0004-6361/202451145},
archivePrefix = {arXiv},
       eprint = {2410.14457},
 primaryClass = {astro-ph.EP},
       adsurl = {https://ui.adsabs.harvard.edu/abs/2024A&A...692A..56Z}
}

@ARTICLE{rosenfeld2014,
       author = {{Rosenfeld}, Katherine A. and {Chiang}, Eugene and {Andrews}, Sean M.},
        title = "{Fast Radial Flows in Transition Disk Holes}",
      journal = {\apj},
         year = 2014,
        month = feb,
       volume = {782},
       number = {2},
          eid = {62},
        pages = {62},
          doi = {10.1088/0004-637X/782/2/62},
archivePrefix = {arXiv},
       eprint = {1312.3817},
 primaryClass = {astro-ph.SR},
       adsurl = {https://ui.adsabs.harvard.edu/abs/2014ApJ...782...62R}
}

@ARTICLE{orihara2025,
       author = {{Orihara}, Ryuta and {Momose}, Munetake},
        title = "{Shadow-based Framework for Estimating Transition Disk Geometries}",
      journal = {\apj},
         year = 2025,
        month = jun,
       volume = {986},
       number = {2},
          eid = {215},
        pages = {215},
          doi = {10.3847/1538-4357/add890},
archivePrefix = {arXiv},
       eprint = {2505.06044},
 primaryClass = {astro-ph.EP},
       adsurl = {https://ui.adsabs.harvard.edu/abs/2025ApJ...986..215O}
}

@ARTICLE{teague2025,
       author = {{Teague}, Richard and {Benisty}, Myriam and {Facchini}, Stefano and {Fukagawa}, Misato and {Pinte}, Christophe and {Andrews}, Sean M. and {Bae}, Jaehan and {Barraza-Alfaro}, Marcelo and {Cataldi}, Gianni and {Cuello}, Nicol{\'a}s and {Curone}, Pietro and {Czekala}, Ian and {Fasano}, Daniele and {Flock}, Mario and {Galloway-Sprietsma}, Maria and {Garg}, Himanshi and {Hall}, Cassandra and {Hammond}, Iain and {Hilder}, Thomas and {Huang}, Jane and {Ilee}, John D. and {Izquierdo}, Andr{\'e}s F. and {Kanagawa}, Kazuhiro and {Lesur}, Geoffroy and {Lodato}, Giuseppe and {Longarini}, Cristiano and {Loomis}, Ryan A. and {Masset}, Fr{\'e}d{\'e}ric and {Menard}, Francois and {Orihara}, Ryuta and {Price}, Daniel J. and {Rosotti}, Giovanni and {Stadler}, Jochen and {Testi}, Leonardo and {Yen}, Hsi-Wei and {Wafflard-Fernandez}, Gaylor and {Wilner}, David J. and {Winter}, Andrew J. and {W{\"o}lfer}, Lisa and {Yoshida}, Tomohiro C. and {Zawadzki}, Brianna},
        title = "{exoALMA. I. Science Goals, Project Design, and Data Products}",
      journal = {\apjl},
         year = 2025,
        month = may,
       volume = {984},
       number = {1},
          eid = {L6},
        pages = {L6},
          doi = {10.3847/2041-8213/adc43b},
archivePrefix = {arXiv},
       eprint = {2504.18688},
 primaryClass = {astro-ph.EP},
       adsurl = {https://ui.adsabs.harvard.edu/abs/2025ApJ...984L...6T}
}

@ARTICLE{kawahara2020,
       author = {{Kawahara}, Hajime and {Masuda}, Kento},
        title = "{Bayesian Dynamic Mapping of an Exo-Earth from Photometric Variability}",
      journal = {\apj},
         year = 2020,
        month = sep,
       volume = {900},
       number = {1},
          eid = {48},
        pages = {48},
          doi = {10.3847/1538-4357/aba95e},
archivePrefix = {arXiv},
       eprint = {2007.13096},
 primaryClass = {astro-ph.EP},
       adsurl = {https://ui.adsabs.harvard.edu/abs/2020ApJ...900...48K}
}

@ARTICLE{longarini2025,
       author = {{Longarini}, Cristiano and {Lodato}, Giuseppe and {Rosotti}, Giovanni and {Andrews}, Sean and {Winter}, Andrew and {Stadler}, Jochen and {Izquierdo}, Andr{\'e}s and {Galloway-Sprietsma}, Maria and {Facchini}, Stefano and {Curone}, Pietro and {Benisty}, Myriam and {Teague}, Richard and {Bae}, Jaehan and {Barraza-Alfaro}, Marcelo and {Cataldi}, Gianni and {Czekala}, Ian and {Cuello}, Nicol{\'a}s and {Fasano}, Daniele and {Flock}, Mario and {Fukagawa}, Misato and {Garg}, Himanshi and {Hall}, Cassandra and {Hammond}, Iain and {Hardiman}, Caitlyn and {Hilder}, Thomas and {Huang}, Jane and {Ilee}, John D. and {Isella}, Andrea and {Kanagawa}, Kazuhiro and {Lesur}, Geoffroy and {Loomis}, Ryan A. and {M{\'e}nard}, Francois and {Orihara}, Ryuta and {Pinte}, Christophe and {Price}, Daniel and {Testi}, Leonardo and {Fernandez}, Gaylor Wafflard- and {W{\"o}lfer}, Lisa and {Yen}, Hsi-Wei and {Yoshida}, Tomohiro C. and {Zawadzki}, Brianna},
        title = "{exoALMA. XII. Weighing and Sizing exoALMA Disks with Rotation Curve Modelling}",
      journal = {\apjl},
         year = 2025,
        month = may,
       volume = {984},
       number = {1},
          eid = {L17},
        pages = {L17},
          doi = {10.3847/2041-8213/adc431},
archivePrefix = {arXiv},
       eprint = {2504.18726},
 primaryClass = {astro-ph.EP},
       adsurl = {https://ui.adsabs.harvard.edu/abs/2025ApJ...984L..17L}
}

@ARTICLE{stadler2025,
       author = {{Stadler}, Jochen and {Benisty}, Myriam and {Winter}, Andrew J. and {Izquierdo}, Andr{\'e}s F. and {Longarini}, Cristiano and {Galloway-Sprietsma}, Maria and {Curone}, Pietro and {Andrews}, Sean M. and {Bae}, Jaehan and {Facchini}, Stefano and {Rosotti}, Giovanni and {Teague}, Richard and {Barraza-Alfaro}, Marcelo and {Cataldi}, Gianni and {Cuello}, Nicol{\'a}s and {Czekala}, Ian and {Fasano}, Daniele and {Flock}, Mario and {Fukagawa}, Misato and {Garg}, Himanshi and {Hall}, Cassandra and {Hammond}, Iain and {Hilder}, Thomas and {Huang}, Jane and {Ilee}, John D. and {Kanagawa}, Kazuhiro and {Lesur}, Geoffroy and {Lodato}, Giuseppe and {Loomis}, Ryan A. and {Menard}, Francois and {Orihara}, Ryuta and {Pinte}, Christophe and {Price}, Daniel J. and {Yen}, Hsi-Wei and {Wafflard-Fernandez}, Gaylor and {Wilner}, David J. and {W{\"o}lfer}, Lisa and {Yoshida}, Tomohiro C. and {Zawadzki}, Brianna},
        title = "{exoALMA. VI. Rotating under Pressure: Rotation Curves, Azimuthal Velocity Substructures, and Gas Pressure Variations}",
      journal = {\apjl},
         year = 2025,
        month = may,
       volume = {984},
       number = {1},
          eid = {L11},
        pages = {L11},
          doi = {10.3847/2041-8213/adb152},
archivePrefix = {arXiv},
       eprint = {2504.20036},
 primaryClass = {astro-ph.EP},
       adsurl = {https://ui.adsabs.harvard.edu/abs/2025ApJ...984L..11S}
}

@article{winter2025,
  author    = {Winter, A.\ J. and Benisty, M. and Izquierdo, A.\ F. and Lodato, G. and Teague, R. and Kimmig, C.\ N. and Andrews, S.\ M. and Bae, J. and Barraza‑Alfaro, M. and Cuello, N. and Curone, P. and Czekala, I. and Facchini, S. and Fasano, D. and Hall, C. and Hardiman, C. and Hilder, T. and Ilee, J.\ D. and Fukagawa, M. and Longarini, C. and Ménard, F. and Orihara, R. and Pinte, C. and Price, D.\ J. and Rosotti, G. and Stadler, J. and Wilner, D.\ J. and Wölfer, L. and Yen, H.-W. and Yoshida, T.\ C. and Zawadzki, B.},
  title     = {exoALMA. XVIII. Interpreting large scale kinematic structures as moderate warping},
  journal   = {preprint, arXiv:2507.11669},
  year      = {2025},
  bibcode   = {2025arXiv250711669W}
}
\end{CJK*}
\end{document}